\documentclass[useAMS,usenatbib]{mn2e}
\usepackage{amsmath}
\usepackage{amssymb}
\usepackage{graphicx}
\usepackage{euscript}
\graphicspath{{a1_ps_figs/}{a2_ps_figs/}{PS_MODEL/}}
\DeclareSymbolFont{rsfso}{U}{rsfso}{m}{n}
\DeclareSymbolFontAlphabet{\mathscr}{rsfso}

\newcommand{\beq}{\begin{equation}}
\newcommand{\eeq}{\end{equation}}
\newcommand{\bfr}{\mbox{\boldmath ${\rm r}$}}
\newcommand{\bfv}{\mbox{\boldmath $v$}}
\newcommand{\bfX}{\mbox{\boldmath $X$}}
\newcommand{\bfB}{\mbox{\boldmath $B$}}
\newcommand{\bfom}{\mbox{\boldmath $\omega$}}
\newcommand{\bfR}{\mbox{\boldmath ${\cal R}$}}
\newcommand{\bfba}{\mbox{\boldmath $\beta$}}
\newcommand{\bfOmega}{\mbox{\boldmath $\Omega$}}
\newcommand{\bfomega}{\mbox{\boldmath $\omega$}}
\newcommand{\bnabla}{\mbox{\boldmath $\nabla$}}
\newcommand{\cendot}{\mbox{\boldmath $\cdot$}}
\newcommand{\ez}{\mbox{{\boldmath $\hat{e}$}}_{Z}}
\newcommand{\cross}{\mbox{\boldmath $\times$}}
\title[Binary systems: Outflows, maser variabilities and baroclinicity]
{Dynamics of stellar wind in a Roche potential: implications for
(i) outflows \& periodicities relevant to astronomical masers, and
(ii) generation of baroclinicity}
\author[Singh \& Deshpande]{Nishant K. Singh$^{1,2,3,4}$\thanks{E-mail:
nishant@nordita.org (NKS)} and Avinash A. Deshpande$^3$\thanks{E-mail: 
desh@rri.res.in (AAD)}\\
$^{1}$Nordita, KTH Royal Institute of Technology and Stockholm University,
Roslagstullsbacken 23, SE-10691 Stockholm, Sweden\\
$^{2}$Inter--University Centre for Astronomy and Astrophysics, Post Bag 4,
Ganeshkhind, Pune 411 007, India\\
$^{3}$Raman Research Institute, Sadashivanagar, Bangalore 560 080, India\\
$^{4}$Joint Astronomy Programme, Indian Institute
of Science, Bangalore 560 012, India
}
\begin{document}
\pagerange{\pageref{firstpage}--\pageref{lastpage}} \pubyear{2012}

\maketitle

\label{firstpage}

\begin{abstract}
We study the dynamics of stellar wind from one of the bodies
in the binary system, where the other body interacts only
gravitationally. We focus on following three issues:
(i) we explore the origin of observed
periodic variations in maser intensity;
(ii) we address the nature of bipolar molecular outflows; and
(iii) we show generation of baroclinicity in the same model setup.
From direct numerical simulations and further numerical
modelling, we find that the maser
intensity along a given line of sight
varies periodically due to periodic modulation of material
density. This modulation period is of the order of the
binary period. Another feature of this model is that the
velocity structure of the flow remains
unchanged with time in late stages of wind evolution. Therefore
the location of the masing spot along the chosen sightline stays at
the same spatial location, thus naturally explaining the observational fact.
This also gives an appearance of bipolar nature
in the standard position-velocity diagram, as has been observed in
a number of molecular outflows.
Remarkably, we also find the generation of baroclinicity
in the flow around binary system, offering
another site where the seed magnetic fields could possibly be generated
due to the Biermann battery mechanisms, within galaxies.
\end{abstract}


\section{Introduction}

Periodic variabilities in the maser intensities have now been observed
in a number of sources, with periods ranging from few days to several
years \citep{GGW04,GMGW05,GLGW09,WGG09,Ara10,SWBL11}.
Such variabilities appear
to be charactersitic of the methanol masers, which trace the early evolutionary
stages of massive star formation \citep{Ell06}.
Massive stars form in regions which are deeply embedded in the
molecular clouds, thus posing many observational challenges. Methanol
masers provide direct access to these regions and therefore prove to
be important tool to study the star formation mechanisms, while also
giving valuable informations on variable conditions in the environment where
these massive stars are born.
This provides sufficient motivation to understand the cause of maser
variabilities, revealed by the long time monitoring of the methanol
masers.

\cite{Nor98} have found from their high-resolution imaging that
some of these sources show arc-like maser spots with velocity
gradients close to those given by Keplerian profile, and thus they argue
that the methanol masers could be possible tracers of circumstellar discs. 
But we also note that there are some sources which do not have such
linear or arc-like structures \citep{Wal98}.
It has also been suggested by \cite{MBC02} that some of these methanol masers
might be associated with molecular outflows or expanding H\,{\small II}
regions, and thus might originate from regions spatially far from those of the
stellar objects.

Much of our understanding of the regions where the young stellar
objects (YSOs) form in the molecular clouds is due to the maser emission from
such locations. High velocity outflows with velocities ranging from 
few $\rm km \,s^{-1}$ to few hundred $\rm km \,s^{-1}$ have been
observed, and it is generally agreed that such outflows occur around
YSOs, driven due to strong stellar winds in the early evolutionary
stages (see reviews \cite{Sne83, Lad85, WVPWB85, SAL87, Bac96}).
These outflows are mostly bipolar and are ubiquitious.
Since the discovery of these bipolar outflows \citep{SLP80} various 
attempts have been made to understand the physical nature of these 
phenomena. 
It is generally believed that the collimation has to 
happen very near to the star forming region and it seems unlikely 
that it happens either due 
to anisotropic environment far away from the central region or also 
due to local direction of magnetic fields 
\citep{CRBC81, Kon82, Tor_etal83, Hey_etal86}.

Given the widely accepted view that most stars are part of multiple
star systems, mostly binaries, and the compelling evidence from
various maser observations that the phenomena, such as, the maser
variabilities, or the collimation of outflows, should occur locally,
we find it important to study the dynamics of stellar wind in the
binary potential.
This also leads us to an important question concerning the generation
of seed magnetic field from initially zero magnetic field, and
we try to study whether such a possibility exists in the same model
setup.
The standard paradigm to explain the orgin of cosmic magnetism involves,
first, the generation of seed magnetic fields, which is later amplified
due to the turbulent dynamos \citep{BS05,Sub08}.
The presence of baroclinicity is known to give rise to
the vorticity, or the magnetic field in the electrically conducting plasma,
starting from zero initial fields (see e.g. \cite{Sub08,Mod14}).
Here, our interest is to only explore if the baroclinicity develops in
our model system, and any furthur studies focussing on the vorticity
or the magnetic fields will be taken up elsewhere.

In the present work, by considering stellar wind from one of the bodies
in the binary system, we focus on following thee issues:
(i) the periodic variations in maser intensity;
(ii) bipolar outflows; and
(iii) the generation of baroclinicity. 
In \textsection~2 we describe our model setup and derive some
general principles in the rotating reference frame.
Results from direct numerical simulations (DNS) are presented and discussed
in \textsection~3. Based on our DNS results, we perform further
simulations to study maser variability and the issue of bipolar outflows
in \textsection~4. We conclude in \textsection~5.

\section{The Model}
We present a simple model which involves basics of the two-body
and the three-body problems in classical mechanics. We repeat some 
of the well known things for completeness and the details can be found
in any standard textbook on classical mechanics (see e. g. 
\cite{VK05,Mor08}).

Consider a \emph{binary system} consisting of two bodies, $S$ and $P$,
which are rotating around their common center of mass, $O$, in a plane.
Let $\overline{X}\,\overline{Y}\,\overline{Z}$ be the \emph{inertial}
(fixed) coordinate frame in which the two bodies lie in the
$\overline{X}\,\overline{Y}-$plane with angular velocity vector in
$\overline{Z}-$direction. Let $X\,Y\,Z$ be the \emph{rotating} (comoving)
coordinate frame which rotates with an angular velocity same as that
of the two bodies in the binary system and therefore both the bodies 
appear to be at rest in this frame. The origins of both the coordinate 
frames coincide and are taken to be at the center of mass, $O$, of the
binary system. All the assumptions made while studying this problem
analytically are given below, but it should be noted that while performing
numerical simulations for this problem, most of the following assumptions
are lifted and therefore we simulate a more realistic case.
\begin{enumerate}
 \item One of the bodies, $S$, has a spherically symmetric wind very near 
to its upper atmosphere, whereas the other body, $P$, is interacting only
gravitationally.
\item $S$ and $P$, which are called the primaries, move in circular orbits
around their common center of mass, $O$, in a plane.
\item The mass of the wind is assumed to be negligible as compared to the
total mass of $S$ and $P$. Thus, it is essentially a \emph{restricted 
circular three-body problem}, with only difference that the wind (massless),
which mimics the third body, is modelled as a \emph{continuum}.
\item Molecular viscosity of the wind is assumed to be negligible.
\item The wind flow is assumed to be in the steady state.
\item We assume that the fluid behaves as a perfect gas and the flow is
\emph{isentropic}. We use perfect gas equation of state for our
analysis.
\end{enumerate}

\begin{figure}
 \includegraphics[scale=0.65]{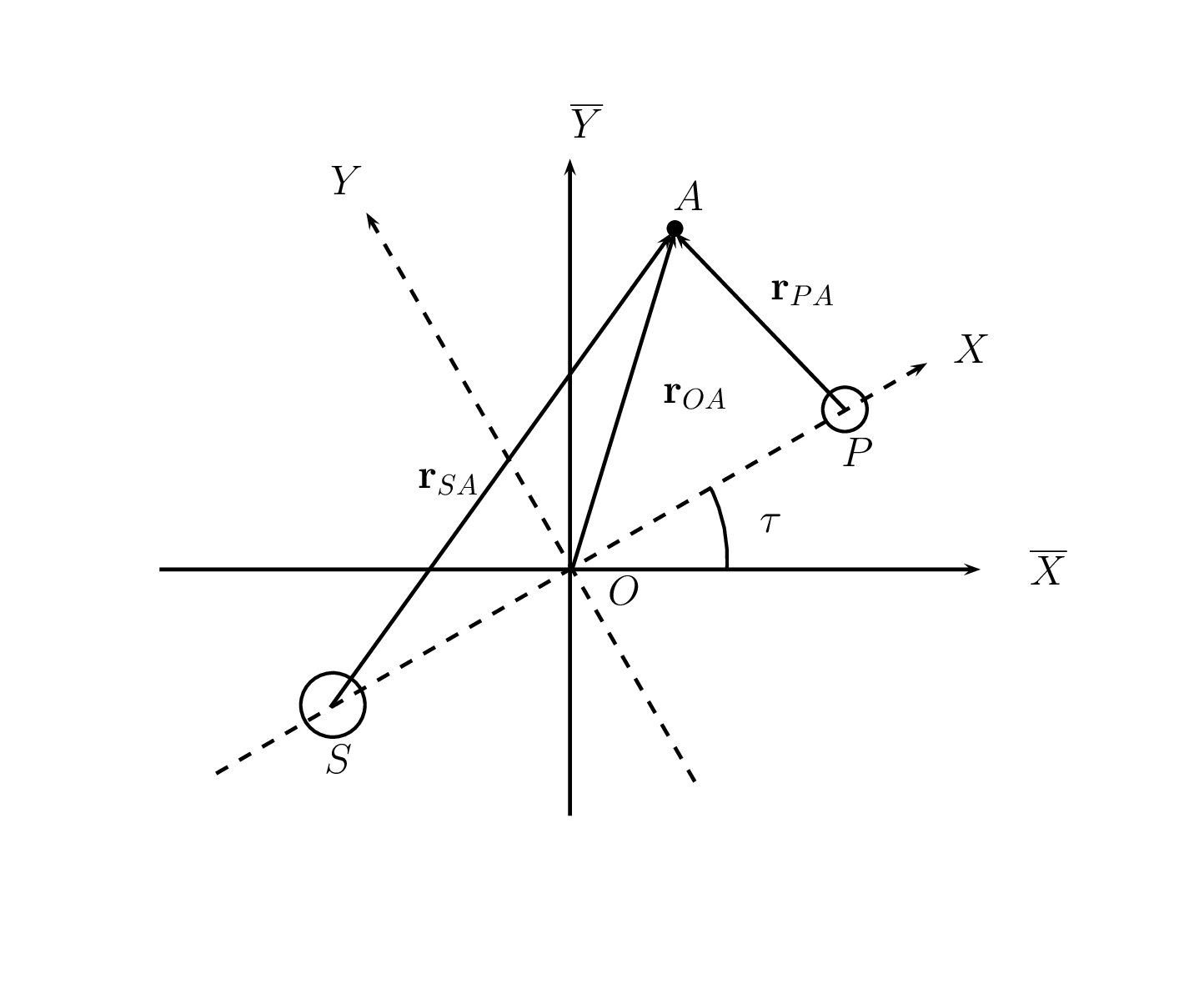}
\caption{The inertial ($\overline{X}\overline{Y}$) and rotating 
($XY$) coordinate frames are shown for the three-body problem.
The center of mass of the two bodies labelled as $S$ and $P$ is the
origin of both the coordinate frames and is denoted by $O$.}
\end{figure}

\noindent
The units of various quantities are chosen such that the properties 
of the system depend only on a single parameter. Let the total mass 
(${\mathscr M}$) of
the primaries ($S$ and $P$) be the unit of mass; the distance between 
them (${\cal D}$) be the unit of distance; and the unit of time be 
chosen in such a way that the angular speed of the primaries, 
denoted by $\Omega$, be unity. It is known that the two bodies circling 
around each other satisfy the following relation:
\beq
\Omega^2 \;=\; \frac{G\,{\mathscr M}}{{\cal D}^3}\;,
\label{kep3}
\eeq    
where $G$ is Newton's gravitational constant.
We find it useful to express Eq.~(\ref{kep3}) in the following form:
\beq
\left(\frac{{\mathscr T}}{1\, {\rm Day}}\right)^2 =
1.33376 \times 10^5\,\left(\frac{{\rm M}_{\odot}}{\mathscr M}\right)
\left(\frac{{\cal D}}{1\, {\rm AU}}\right)^3
\label{kep3-2}
\eeq
where ${\mathscr T}=2\pi/\Omega$ is the time-period of the binary and
${\rm M}_{\odot}$ represents the solar mass.

\begin{figure}
\includegraphics[width=\columnwidth]{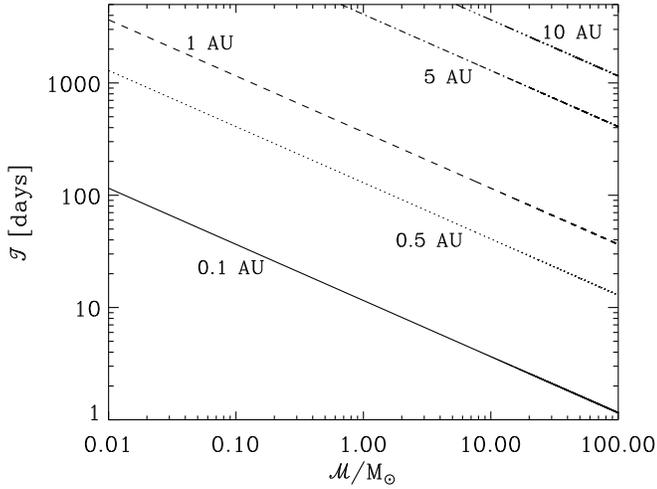}
\caption{
Relation between the total mass (${\mathscr M}$) and the time-period
(${\mathscr T}$) of the binary. Different lines in this log-log plot
correspond to different values of the binary separation (${\cal D}$).
}
\label{mp}
\end{figure}

In dimensionless units, 
equation~(\ref{kep3}) implies $G = 1$ due to the choices made above for 
the units of mass, length and time; also, the mean anomaly equals time 
($\tau = \overline{\tau}$ is the time as seen in both the reference frames).
Figure~1 shows the binary system at an arbitrary time $\tau$ as seen in 
both the coordinate systems, in which $A$ is an arbitrary point, and 
$\bfr_{SA}$, $\bfr_{PA}$ and $\bfr_{OA}$ are the radius vectors of 
point $A$ with respect to $S$, $P$ and $O$ respectively (the axis 
$Z = \overline{Z}$ is not shown explicitly). Let $\xi$ be the mass of $P$,
thus mass of $S$ is $(1 - \xi)$. In the rotating reference frame, with
positive $X$ in the direction of the body $P$, the 
coordinates of $P$ and $S$ will be $(1 - \xi, 0)$ and $(-\xi, 0)$ 
respectively. Let $(X, Y, 0)$ be the coordinate of the arbitrary point 
$A$, assumed to lie in the plane of the binary.
Thus magnitudes of vectors $\bfr_{SA}$ and $\bfr_{PA}$ are given as:
\begin{eqnarray}
 &&{\rm r}_{SA} = \lvert \bfr_{SA} \rvert = \sqrt{(X + \xi)^2 + Y^2} 
\nonumber \\ [2ex]
&&{\rm r}_{PA} = \lvert \bfr_{PA} \rvert = \sqrt{(X - (1 -\xi))^2 + Y^2}
\label{SPcoords}
\end{eqnarray}
\noindent
The gravitational potential in the comoving frame at the point $A$ 
may be written as,
\beq
\Phi \;=\; -\frac{(1 - \xi)}{{\rm r}_{SA}} - \frac{\xi}{{\rm r}_{PA}}
\label{pot}
\eeq

\subsection{Bernoulli's principle in the rotating coordinate system}
If $\bfv(\bfX, \tau)$ be 
the fluid velocity of the wind in the rotating frame
then we may write the Euler equations in rotating frame for steady flow, with
$p$ and $\rho$ as the fluid pressure and density, respectively,
as,
\beq
\left( \bfv \cendot \bnabla \right)\bfv \,=\,-\frac{\bnabla p}{\rho}
-\bnabla \Phi - \hat{\bfOmega} \times \left( \hat{\bfOmega} \times \bfX \right)
- 2\hat{\bfOmega} \times \bfv
\label{Euler1}
\eeq
\noindent
where $(\bfX, \tau) \equiv (X, Y, Z, \tau)$ and $\hat{\bfOmega}$ ($= \ez$,
which is the unit vector along $Z\equiv\overline{Z}$) is the angular velocity
of the comoving frame relative to the inertial frame. Note that the
equation~(\ref{Euler1}) is written in dimensionless form.
Using vector identities, we can write equation~(\ref{Euler1})
in the following form:
\beq
\bnabla {\cal B} + (\bfomega + 2 \hat{\bfOmega})\times \bfv \;=\; {\bf 0}
\label{Euler2}
\eeq
\noindent
where,
\beq
{\cal B} \;=\; \left( \frac{1}{2}v^2 + \int \frac{d p}{\rho} + 
\Phi_{\rm eff} \right)\;;\qquad \bfomega \;=\; \bnabla \times \bfv
\label{bernvort}
\eeq
\noindent
The effective potential ($\Phi_{\rm eff}$) in equation~(\ref{bernvort}) is
written as,
\beq
\Phi_{\rm eff} \;=\; \Phi - \frac{1}{2}\lvert \hat{\bfOmega}\times
\bfX \rvert^2 \;=\; -\frac{(1 - \xi)}{{\rm r}_{SA}} - 
\frac{\xi}{{\rm r}_{PA}} - \frac{1}{2}\lvert \hat{\bfOmega}\times
\bfX \rvert^2
\label{effpot}
\eeq
\noindent
On taking the dot product of equation~(\ref{Euler2}) with $\bfv$, we obtain
\beq
(\bfv \cendot \bnabla)\,{\cal B} \;=\; 0
\label{bernth}
\eeq 
\noindent
Therefore the quantity, ${\cal B}$, is a constant along a particular streamline
for steady flows, although it could be a different constant for different 
streamlines. Noting the fact that the particle paths and streamlines are the 
same for steady flows, we can see that ${\cal B}$ remains the same for a 
particular fluid element as it moves along a particular streamline.

\begin{figure}
\includegraphics[width=\columnwidth]{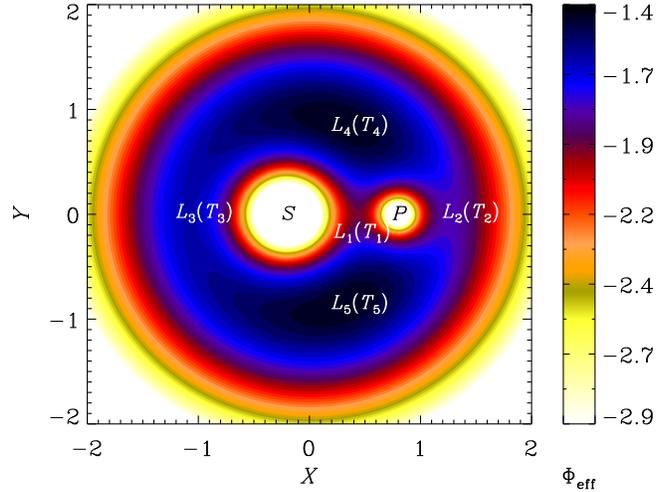}
\caption{Isocontours of the effective potential $\Phi_{\rm eff}$
given by Eq~(\ref{effpot}), for $\xi=0.2$.
Locations of Lagrangian and the temperature points are shown by
$L_1 - L_5$ and $T_1 - T_5$, respectively.}
\end{figure}

\subsection{Nature of the flow}
The Lagrangian time derivative is the rate at which some quantity of interest
changes as we follow any particular fluid element, which is denoted as,
\beq
\frac{d}{d \tau} \;=\; \frac{\partial}{\partial \tau} + 
(\bfv \cendot \bnabla)
\label{lagder}
\eeq
\noindent
As the first term on the right hand side of equation~(\ref{lagder}) 
(which is the Eulerian time derivative) does not
contribute for steady flows, we can write from equations~(\ref{bernth})
and (\ref{lagder}),
\beq
\frac{d {\cal B}}{d \tau} \;=\; (\bfv \cendot \bnabla)\,{\cal B} \;=\; 0
\label{bernth2}
\eeq
\noindent
Equation~(\ref{bernth2}) implies that, for any streamline, which is 
same as the trajectory of a particular fluid element, we may write
\beq
{\cal B} = \frac{1}{2}v^2 + \left(\frac{\gamma}{\gamma - 1}\right) R\, T + 
\Phi_{\rm eff} = \textrm{constant} = C
\label{bernth3}
\eeq
\noindent
where we have used adiabatic equation of state and note that the term
$\int dp / \rho$ appearing in equation~(\ref{bernvort}) may be replaced by
specific enthalpy ($w$) for isentropic evolution of fluid element. We know,
\beq
w \;=\; \left(\frac{\gamma}{\gamma - 1}\right) R\, T
\label{spenth}
\eeq
\noindent
where $\gamma$ is ratio of specific heats at constant pressure and constant 
volume, $R$ is the gas constant and $T$ is the temperature. As the terms
$v^2 / 2$ and $\{\gamma / (\gamma -1)\}\, R T$ in equation~(\ref{bernth3})
cannot be negative, we infer from equation~(\ref{bernth3}) that the motion
of a fluid element, and hence the corresponding streamline, is restricted
to the region where 
\beq
\Phi_{\rm eff} < C
\eeq

\subsection{Isocontours of effective potential and temperature: 
zero velocity curves}

It is known that the isocontours of the effective potential exhibit 
five Lagrangian points, at which the test-body may remain at rest in
the comoving frame. 
These points are extrema or saddle points of $\Phi_{\rm eff}$.
It is remarkable to note that \emph{the shape of the isocontours 
of the temperature distribution for binary system under consideration 
is identical to the isocontours of the effective potential}, as may be 
seen from the following discussion. 

Let us focus on the zero-velocity-curves, the topology of which depends
on the energy of a particular fluid element. Let $\Phi_{\rm eff} = C_1$,
where $C_1$ is constant. Then, we can see from equation~(\ref{bernth3})
for zero-velocity-curves, that
\begin{eqnarray}
\left(\frac{\gamma}{\gamma - 1}\right) R\, T\,&=&\,C - \Phi_{\rm eff}
\,=\,C - C_1 \,=\, \textrm{constant} \nonumber \\ [2ex]
\Rightarrow \quad \quad T \,&=&\, \textrm{constant, where,}\,\, 
\Phi_{\rm eff} = C_1\,,
\label{tempiso}
\end{eqnarray}
\noindent
as $\{\gamma / (\gamma -1)\}\, R$ is a constant for a particular value of 
$\gamma$. Thus we conclude that the shapes of the isocontours of the effective
potential and the temperature are same, whereas the values might be different.
Hence we will have five special \emph{temperature points}, $T_1 - T_5$, 
at the same locations where we have five special \emph{Lagrangian points}, 
$L_1 - L_5$. We may refer to these five temperature points as 
\emph{Lagrangian-equivalent-temperature points}.

\section{Direct numerical simulations}

So far, we have studied the behavior of few scalar variables along 
streamlines. We now wish to know the trajectories of fluid elements, i.e.,
we would like to study how the spherically symmetric wind from the body
$S$ flows in the presence of another gravitating body $P$ in a binary system.
For such investigations, we use PLUTO code 
\footnote{See http://plutocode.ph.unito.it/.}, which is 
a Godunov-type modular
code intended primarily for computational astrophysics and high mach number
flows in multiple spatial dimensions.
The code allows us to study two-dimensional problems in a frame rotating
with constant angular velocity pointing along $Z$-direction, by suitably
adding noninertial (the coriolis and the centrifugal) forces in the momentum
equation.
The details of the code may be found
in \cite{Mig_etal07} (and references therein).
Our strategy to study the dynamics of the wind in a binary system
using PLUTO code may be expressed as follows: 

\begin{enumerate}
\item We adapt the code in the comoving frame of the binary in which the two bodies,
$S$ and $P$, appear to be at rest, by adding the necessary body-forces,
namely, coriolis and centrifugal, to the equation of motion. 

\item We use the hydrodynamic module of the code and solve the
equations in two-dimensional $r$-$\phi$ plane, where $X=r \cos{\phi}$
and $Y=r \sin{\phi}$, and the angular velocity points in the
$Z$-direction. Linearized Roe Riemann solver has been
used for flux computation and we used outflow boundary conditions at
the outer edge of the domain.

\item As we wish to study how the wind from one of the bodies ($S$) in binary
system flows due to purely gravitational effects, we do not consider the effect
of forces that might accelerate the wind radially outwards from $S$ when it 
leaves the surface of the body $S$ (e. g. radiation pressure on the wind due 
to $S$ may lead to radially outward acceleration of the wind). One may wish to 
imagine a radially-outward-acceleration-zone around body $S$, beyond which,
the dynamics of the wind is solely governed by the gravitational effects due to
binary system and the pressure gradients. The radius of such acceleration
zone was chosen to be equal to $20\,\%$ of the binary separation. 

\item Thus we consider the spherically symmetric wind from $S$ and study its
dynamics in the plane of the binary.
Equation of motion for the fluid particles is given by:
\beq
\frac{\partial \bfv}{\partial \tau} \,+\,
\left( \bfv \cendot \bnabla \right)\bfv \,=\,-\frac{\bnabla p}{\rho}
-\bnabla \Phi_{\rm eff} - 2\hat{\bfOmega} \times \bfv
\label{Euler_unsteady}
\eeq 
\noindent
where symbols have ususal meanings described earlier.
As our aim is to illustrate the physical mechanism, we are not
much interested in absolute values of various physical quantities,
but their relative changes as one moves
about in space at any given time, would indeed be useful for our further
modelling.
\end{enumerate}

\begin{figure*}
\includegraphics[scale=0.3]{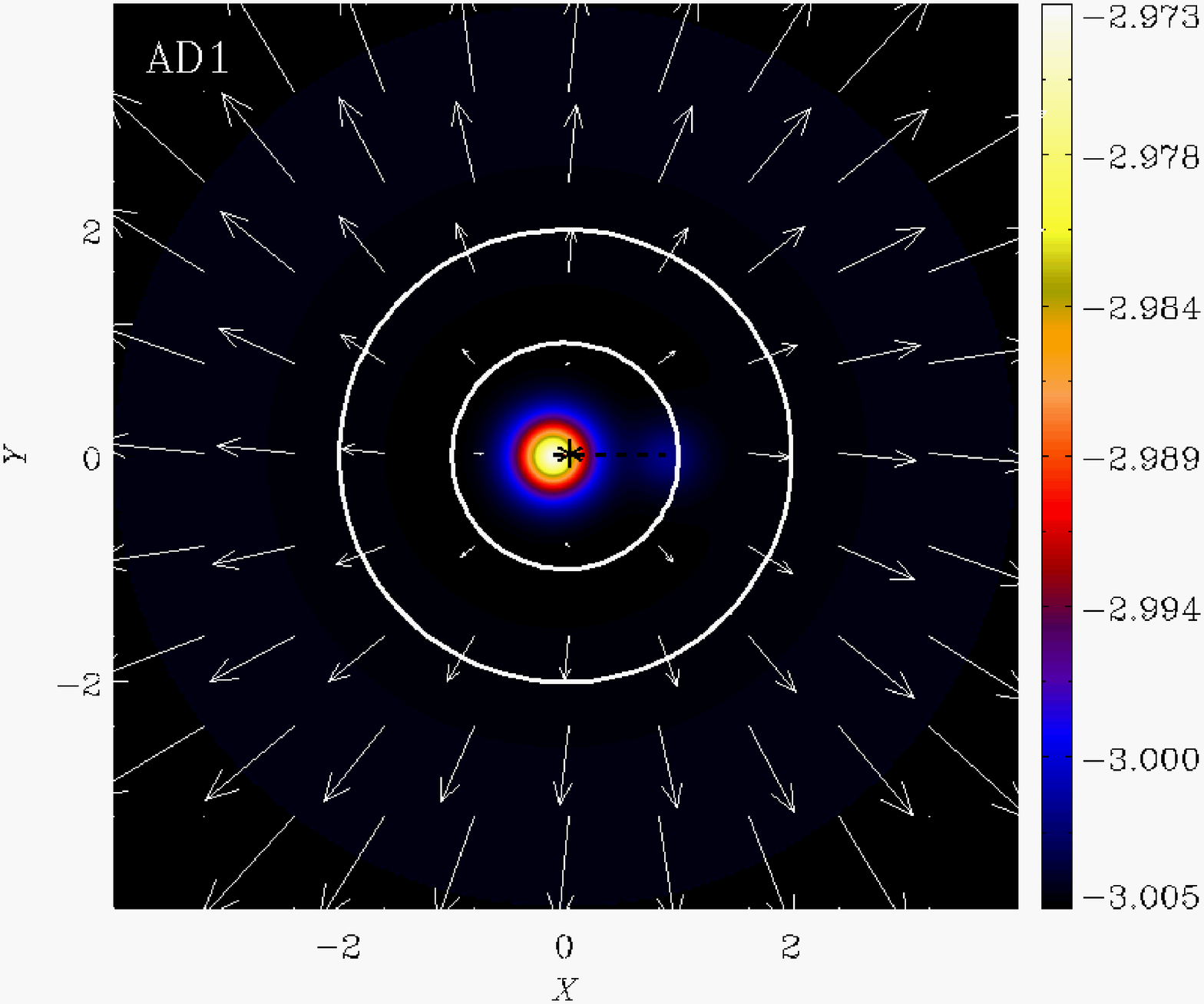}
\includegraphics[scale=0.3]{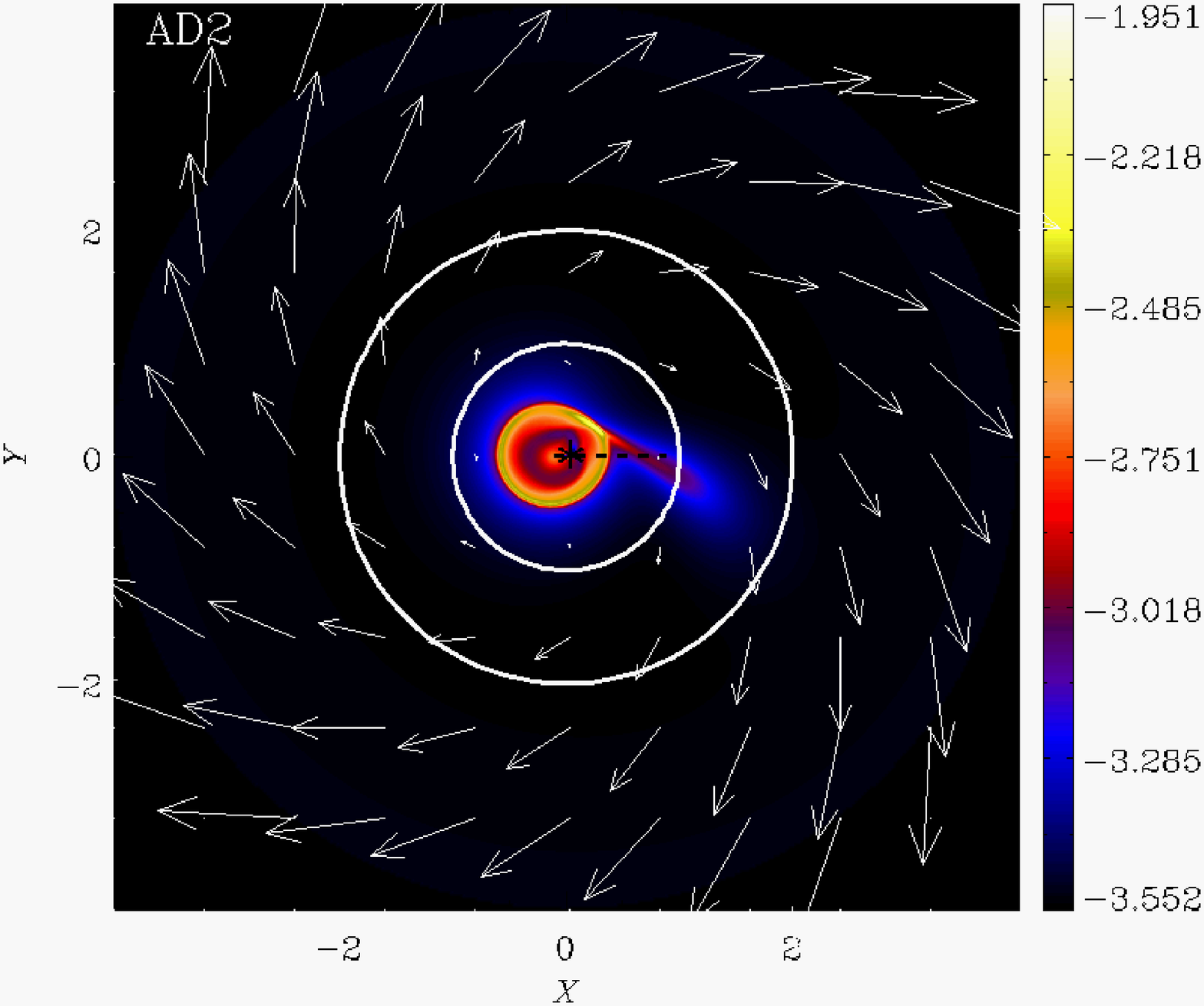}
\vskip0.01in
\includegraphics[scale=0.3]{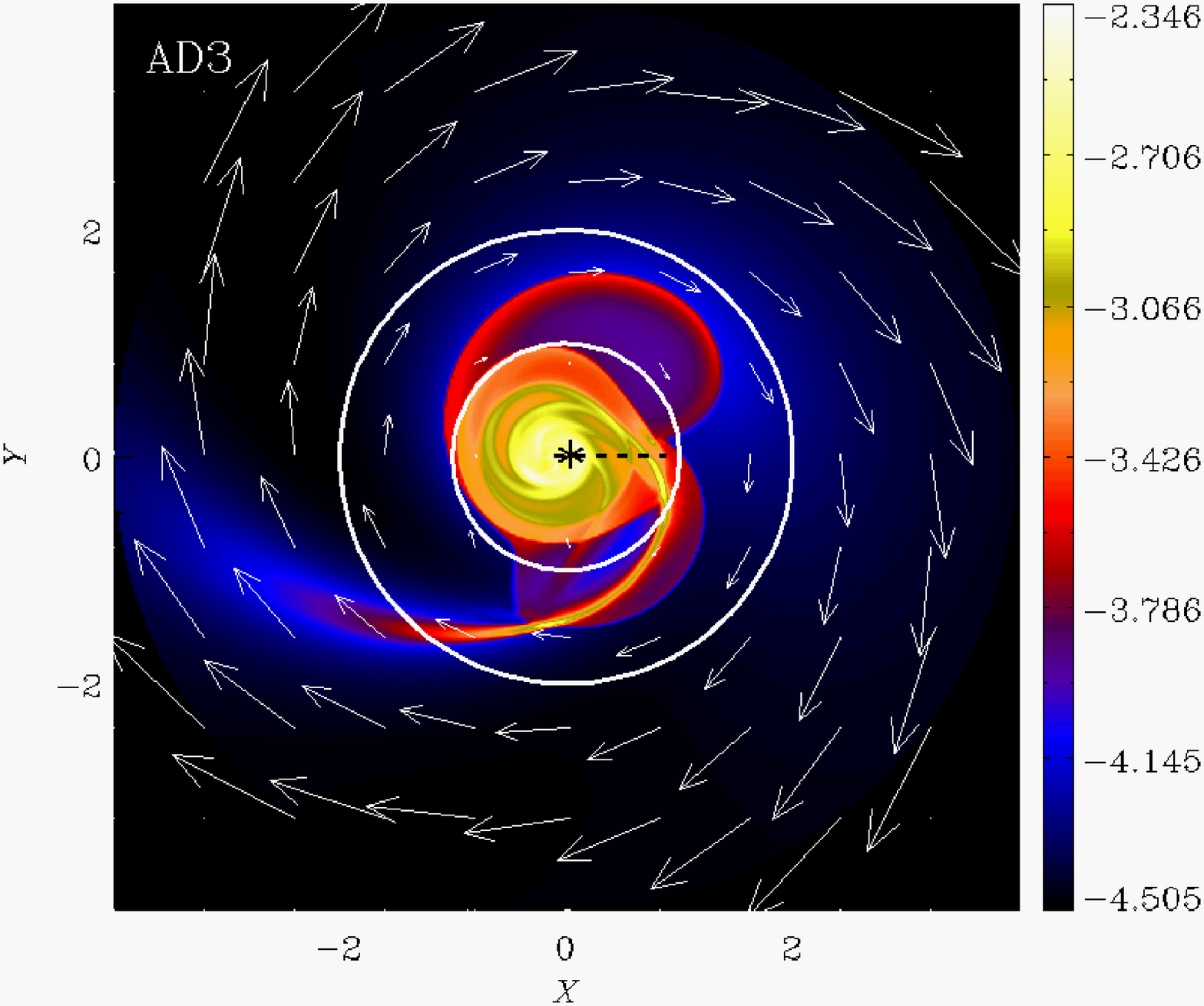}
\includegraphics[scale=0.3]{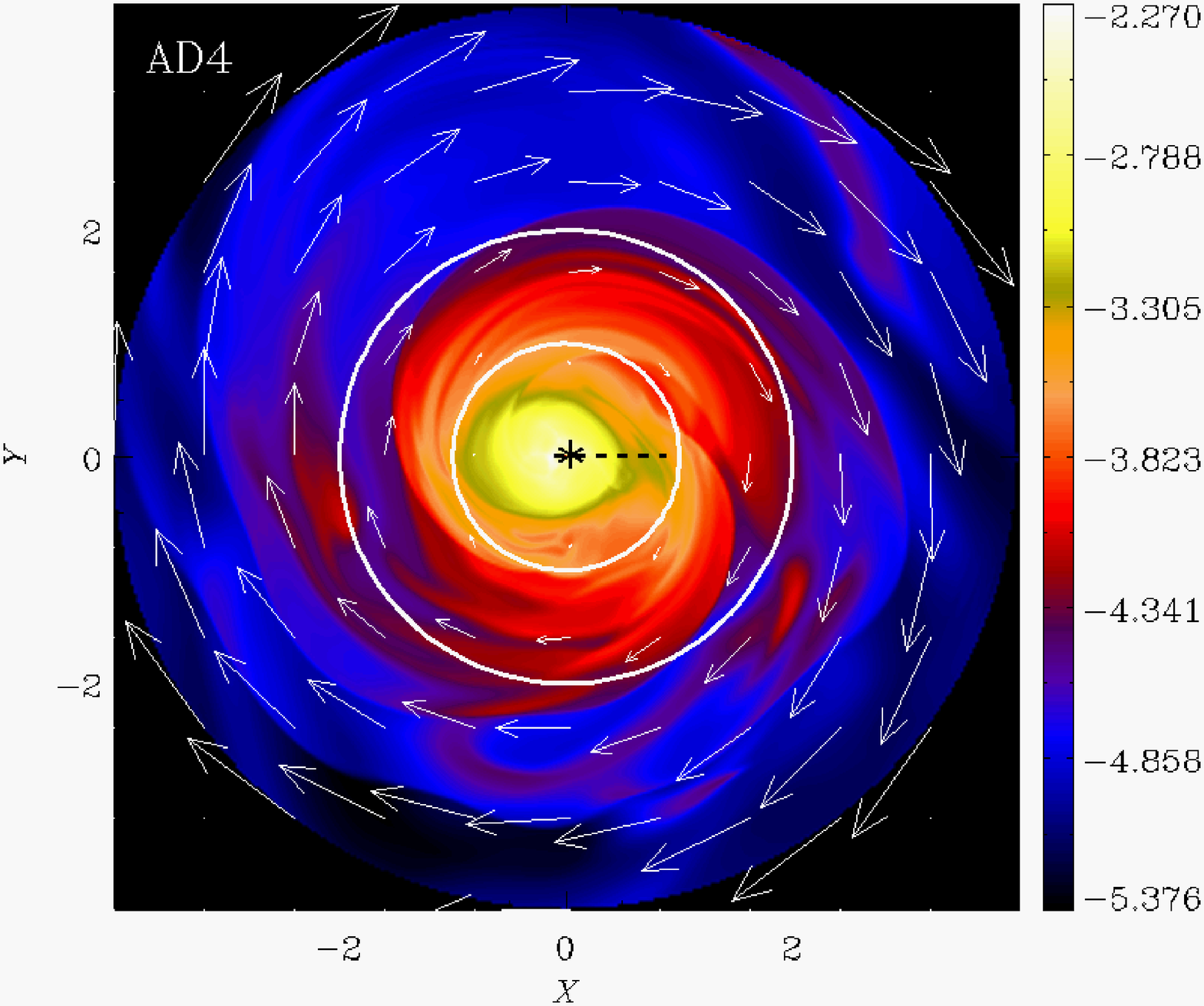}
\vskip0.01in
\includegraphics[scale=0.3]{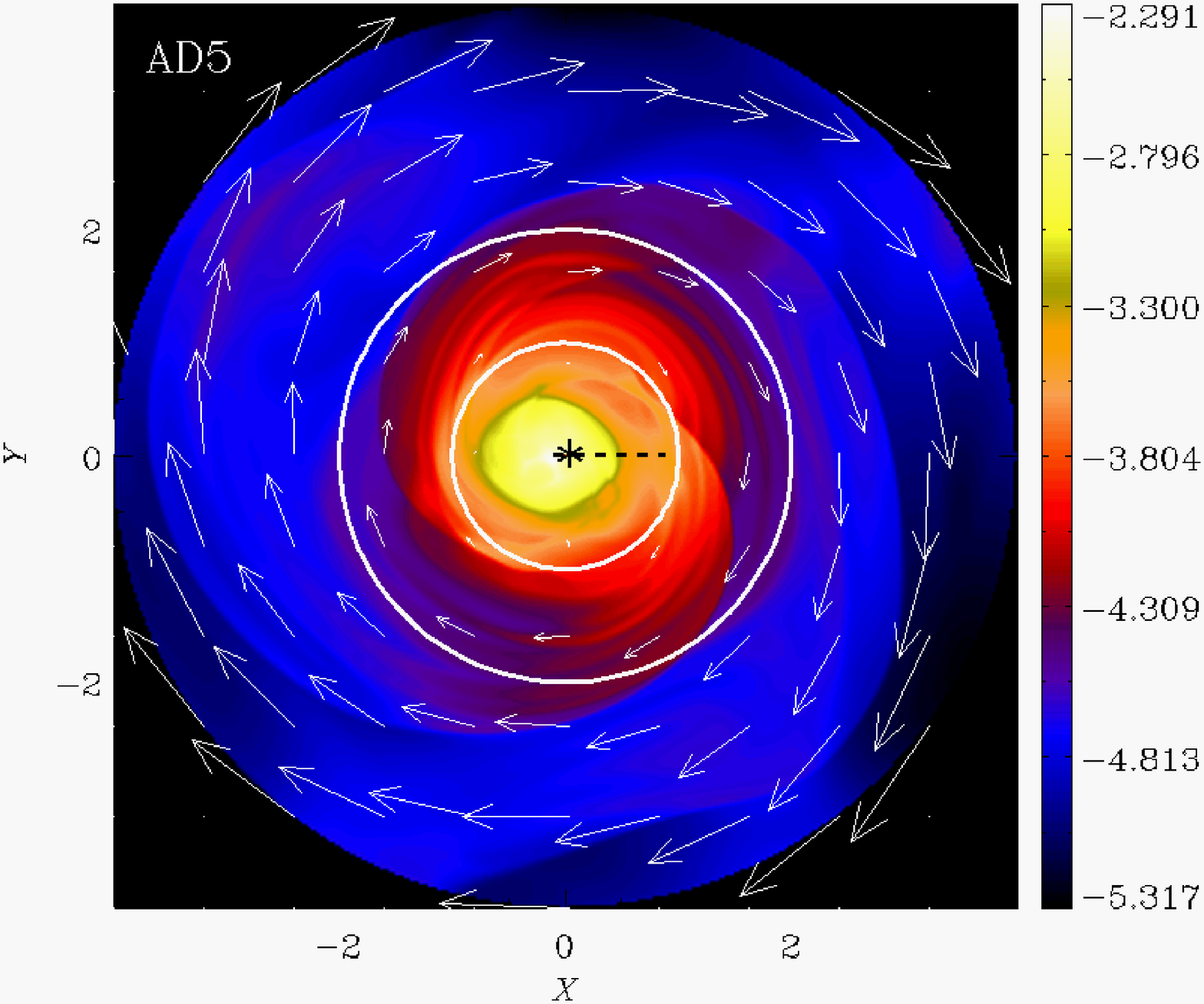}
\includegraphics[scale=0.3]{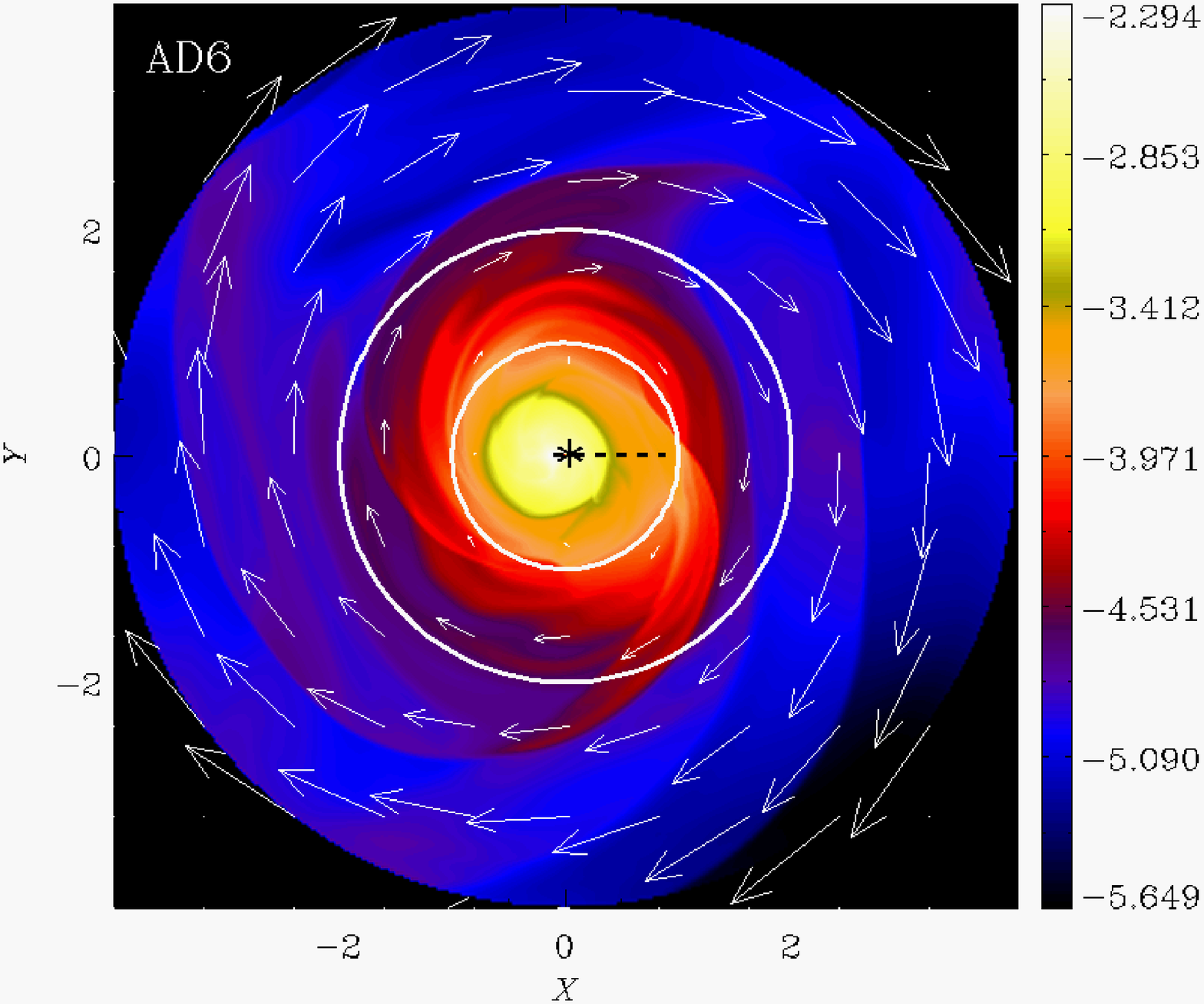}
\caption{
($\xi=0.1$) Snap-shots of logarithmic density maps, as seen from the
corotating frame; colours indicate $\log_{10}\rho$.
Time increases from upper-left to lower-right
panel. Two concentric circles are centered about the common center
of mass of the binary, denoted by the asterisk (`$\ast$'). The dashed
line joins the centers of the two components of the binary, where
the star, $S$ (with the wind), lies at the left end, whereas the other
body, $P$, lies at the right end.
Arrows indicate the velocity vector field.
}
\label{rho_o1}
\end{figure*}

\begin{figure*}
\includegraphics[scale=0.3]{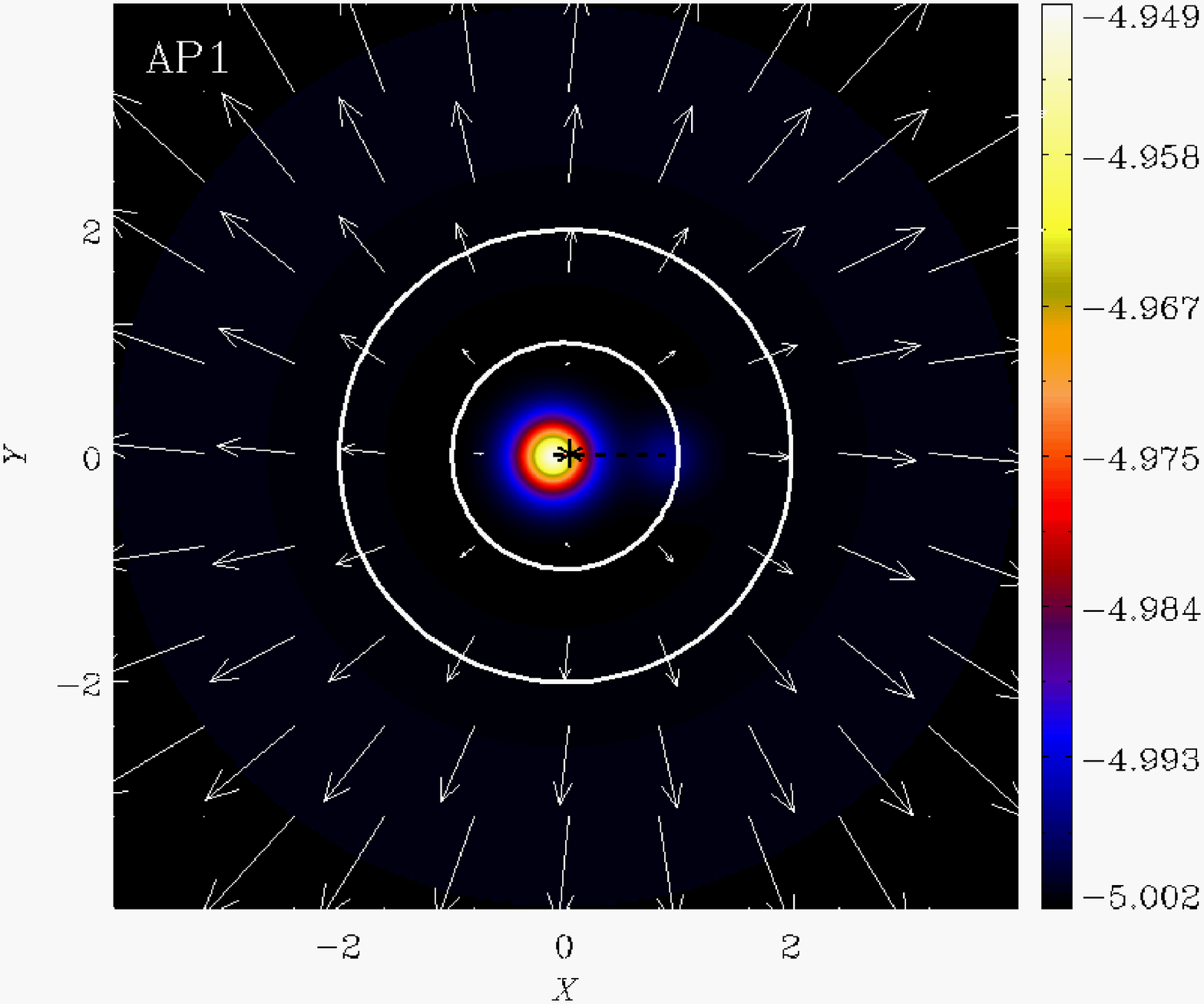}
\includegraphics[scale=0.3]{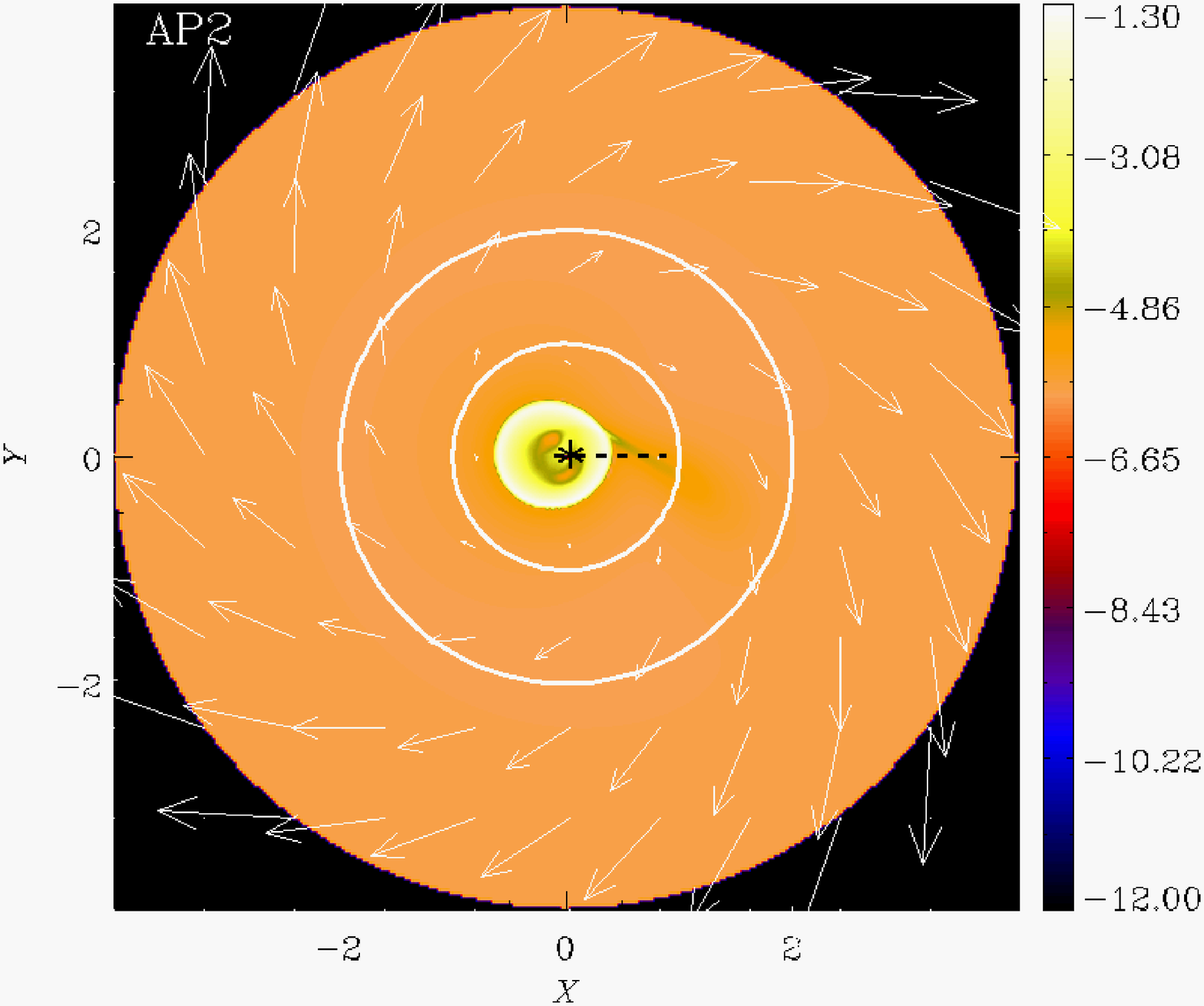}
\vskip0.01in
\includegraphics[scale=0.3]{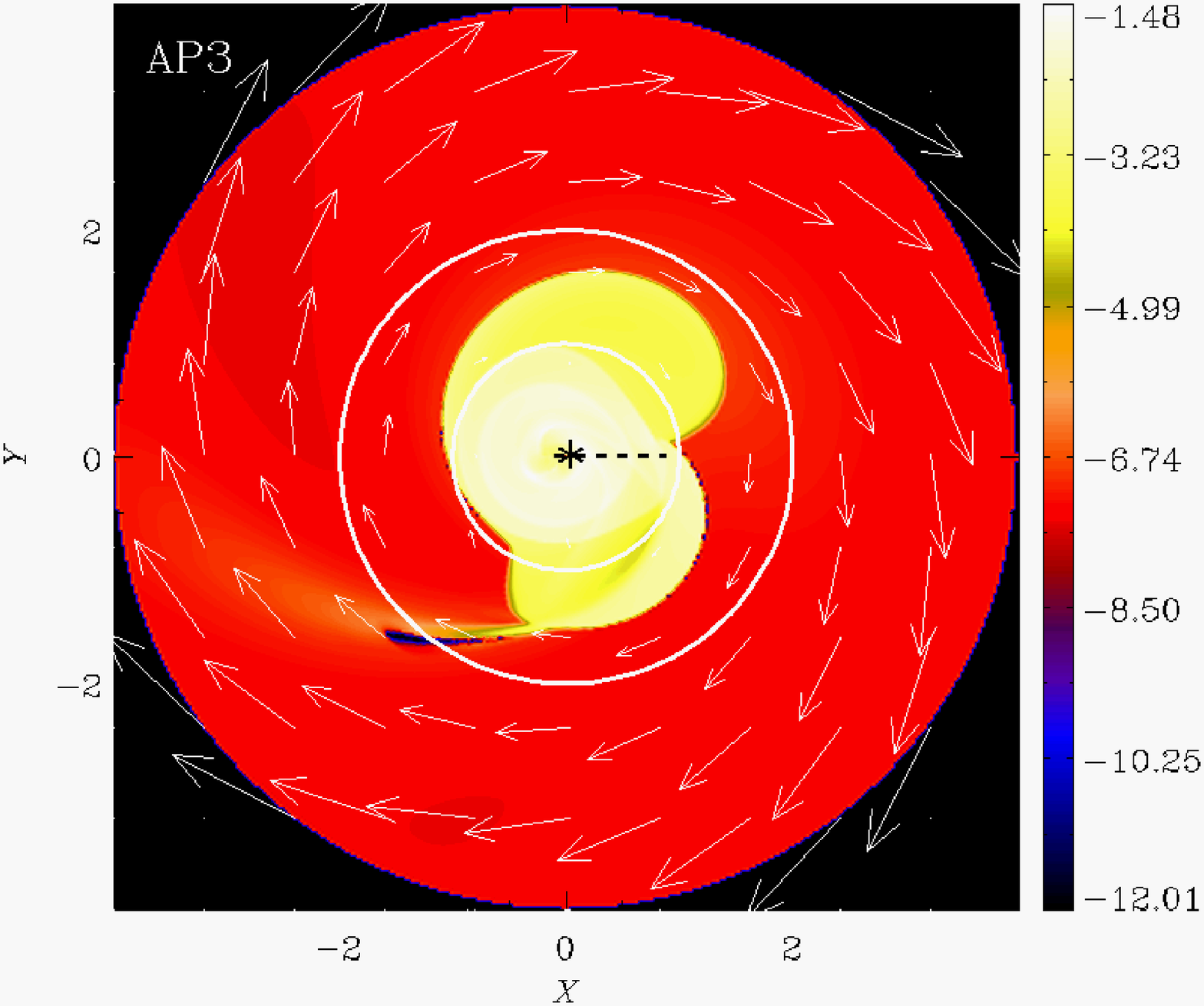}
\includegraphics[scale=0.3]{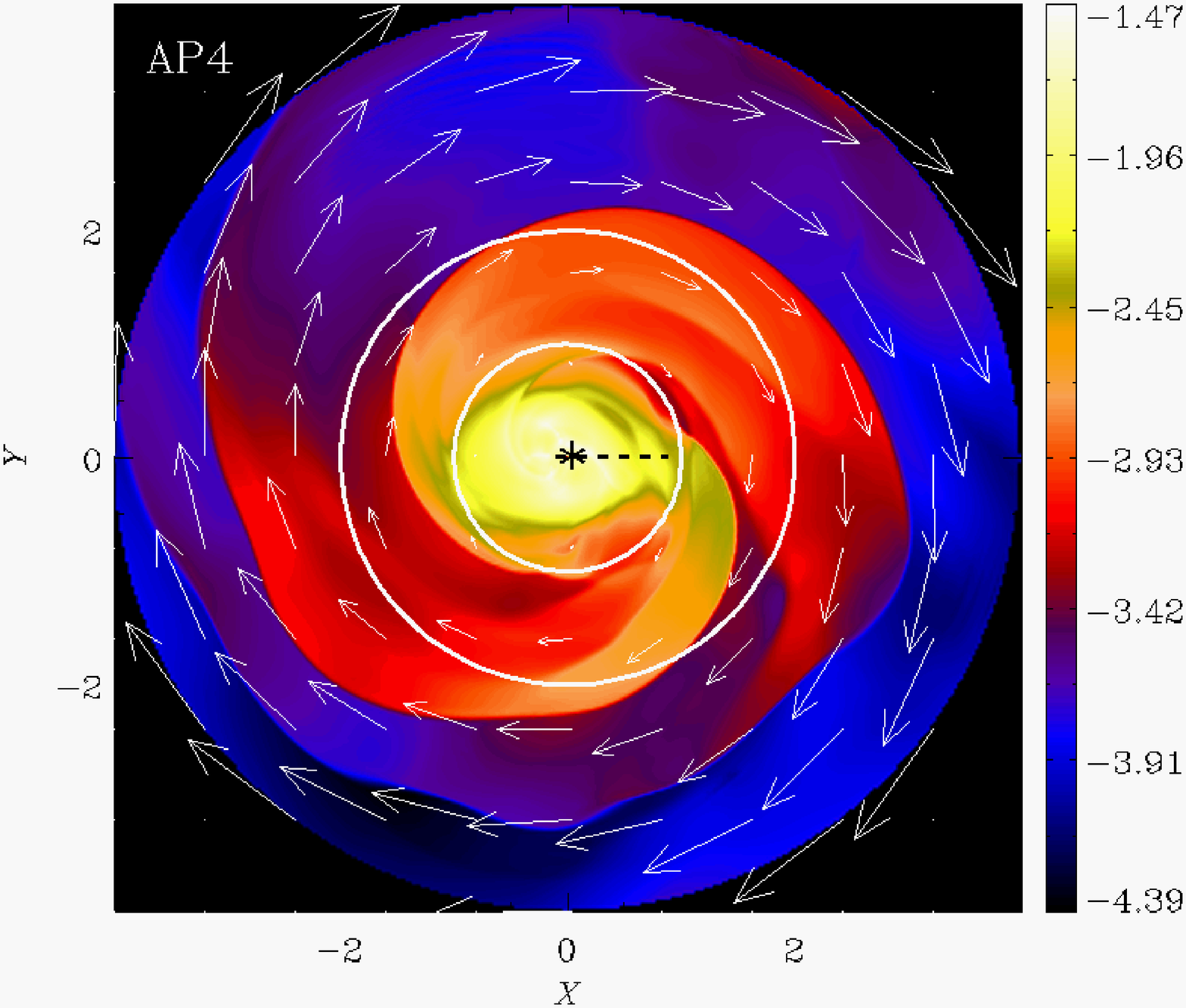}
\vskip0.01in
\includegraphics[scale=0.3]{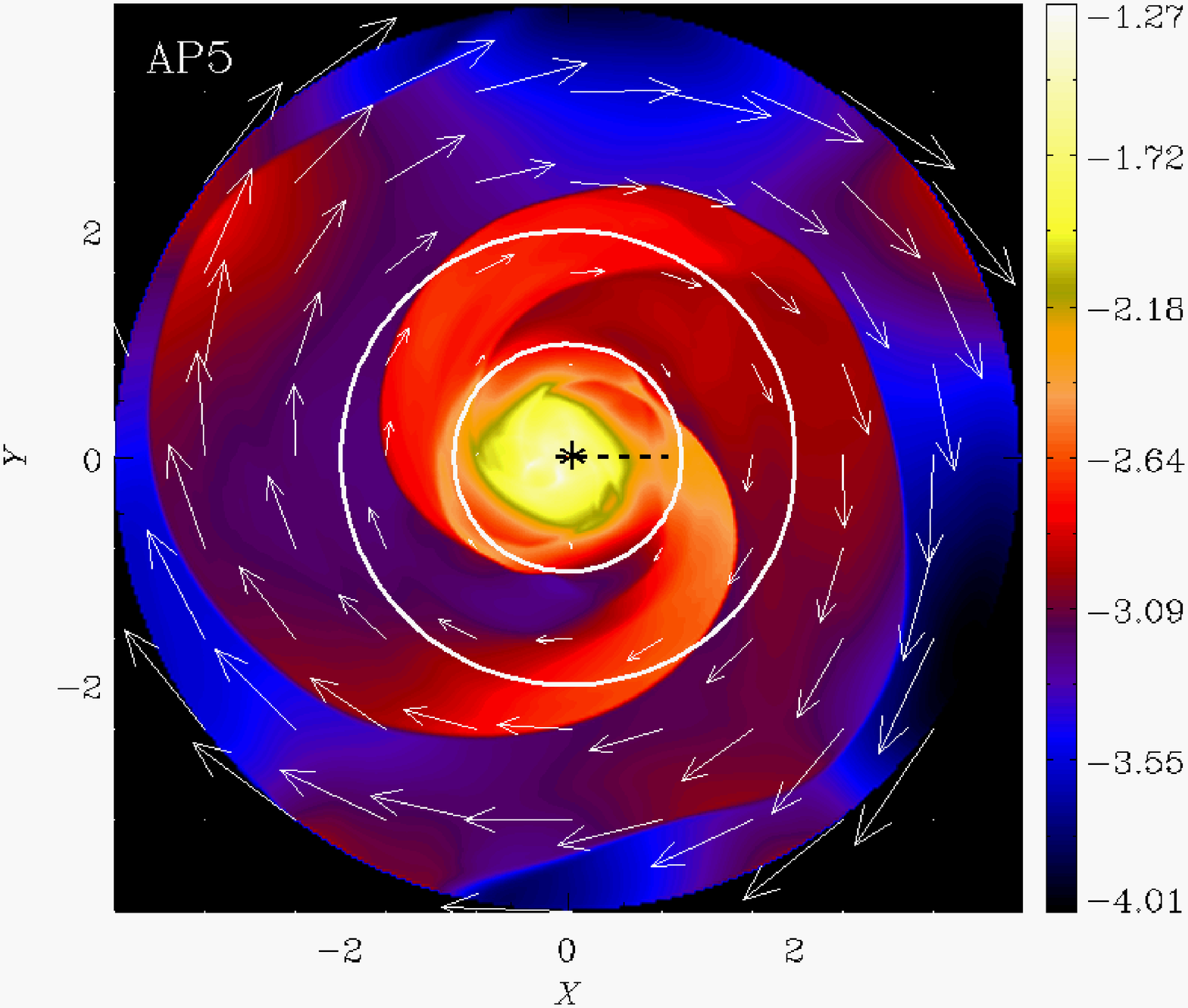}
\includegraphics[scale=0.3]{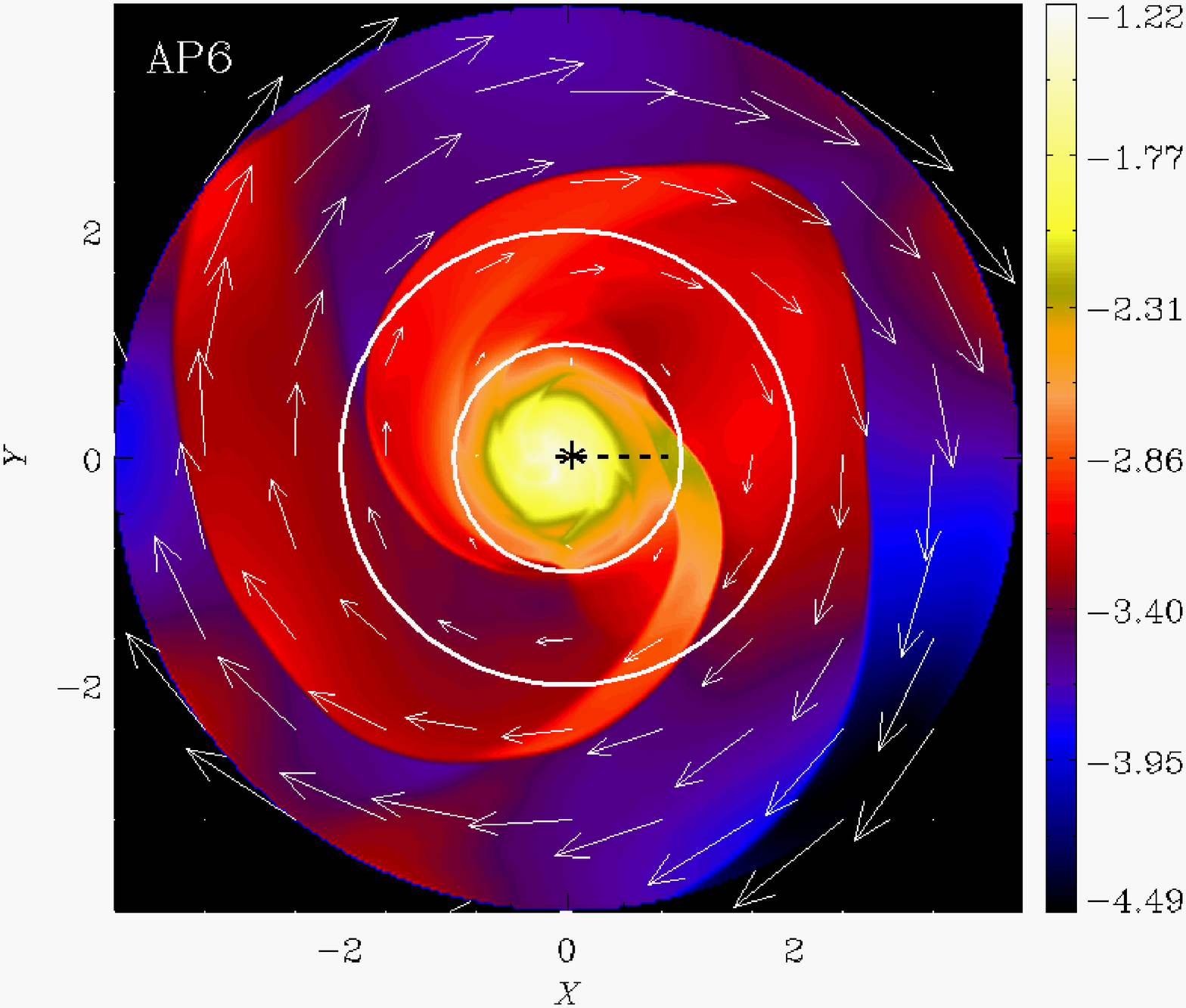}
\caption{
Similar to Fig.~(\ref{rho_o1}) but shows the snap shots of the
pressure maps on logarithmic scale ($\log_{10}P$).
\label{prs_o1}
}
\end{figure*}

\begin{figure*}
\includegraphics[scale=0.3]{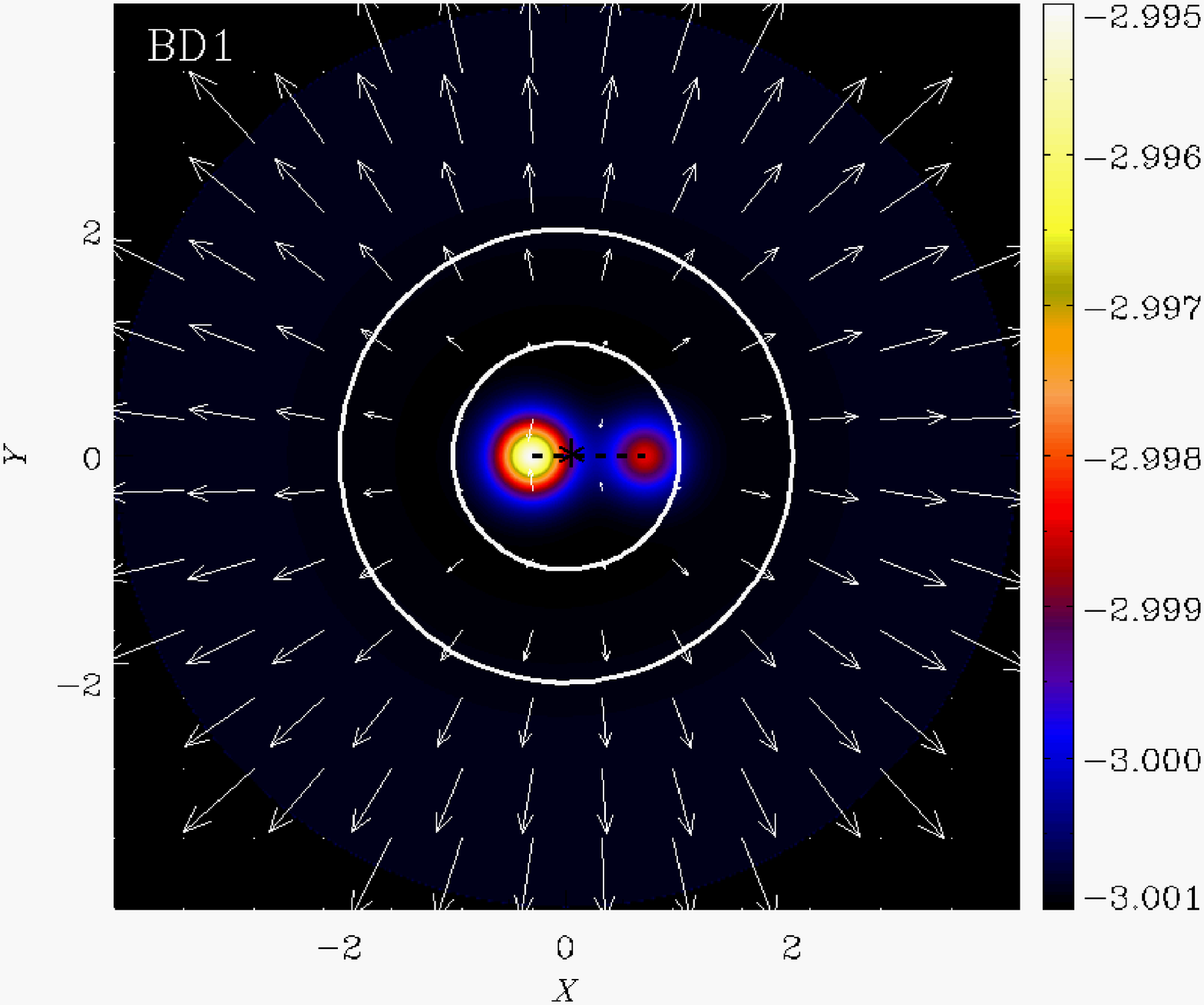}
\includegraphics[scale=0.3]{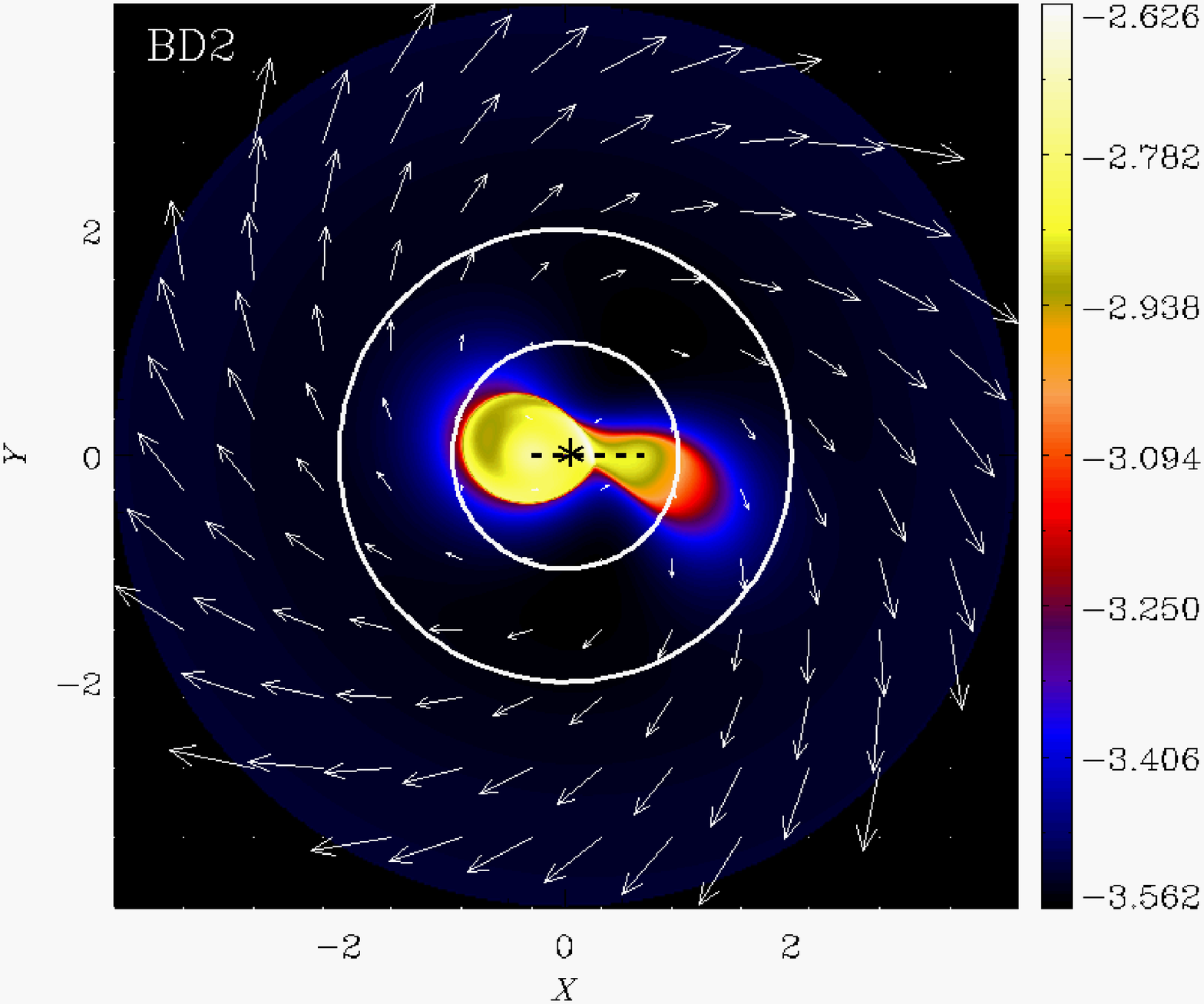}
\vskip0.01in
\includegraphics[scale=0.3]{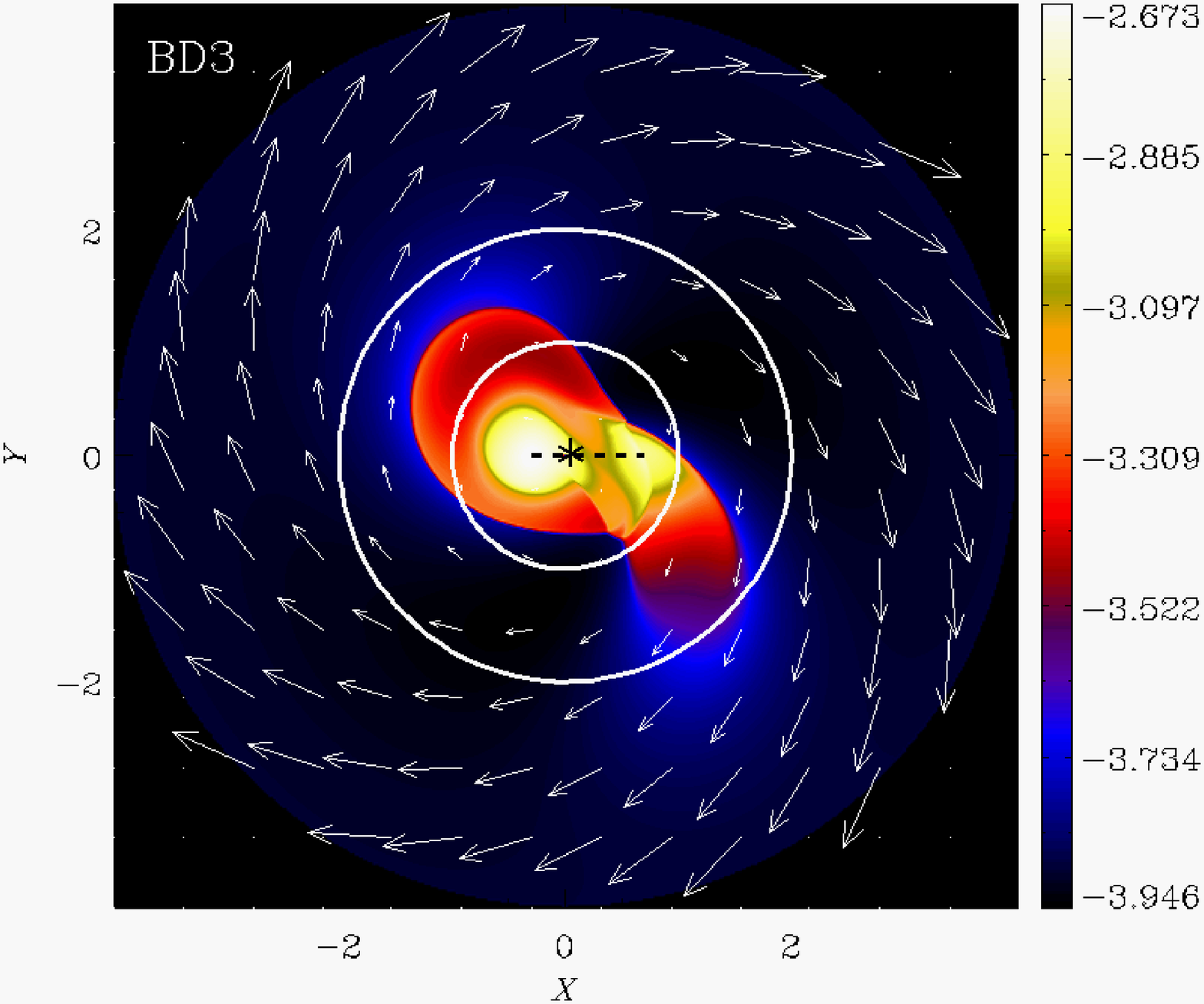}
\includegraphics[scale=0.3]{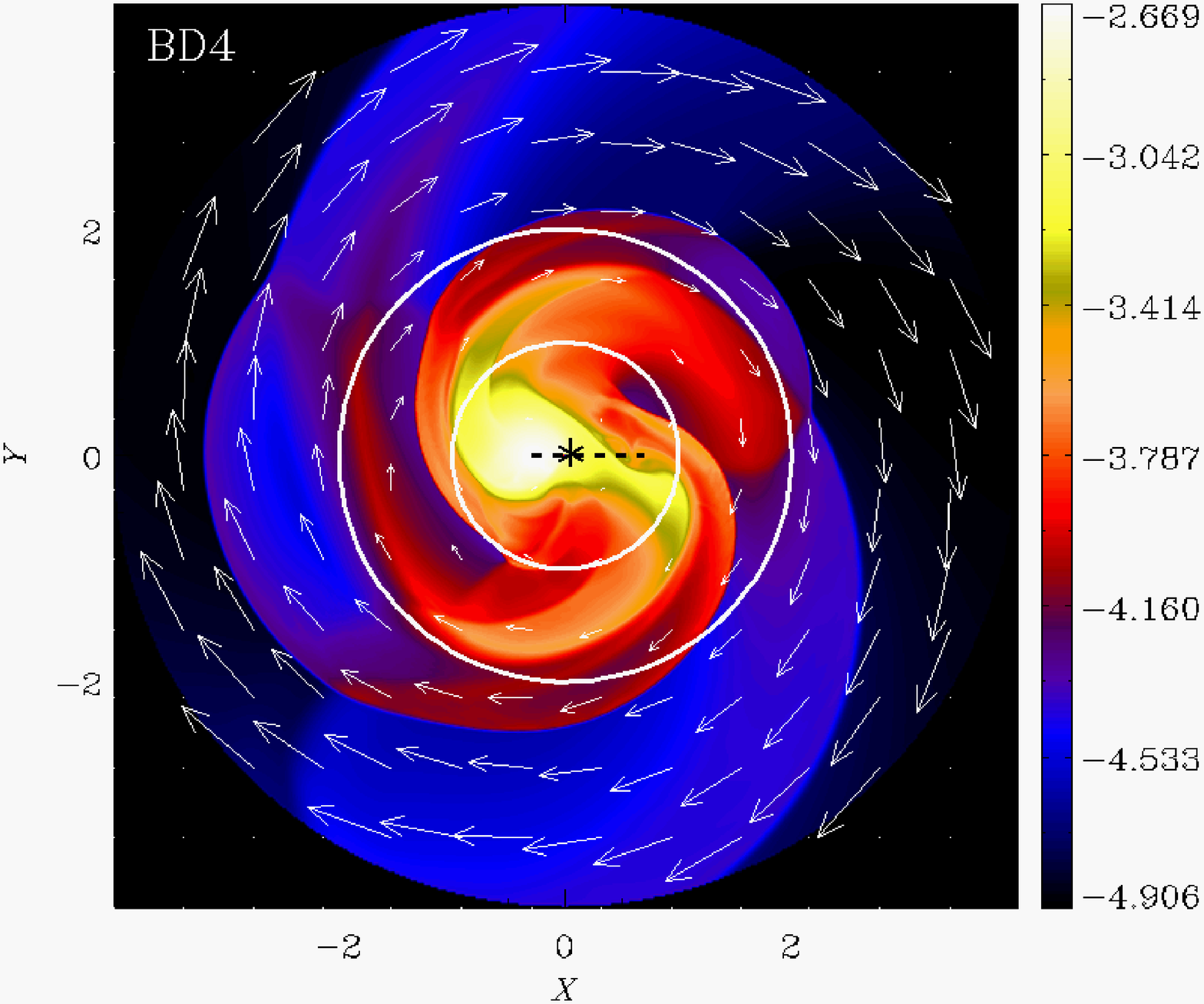}
\vskip0.01in
\includegraphics[scale=0.3]{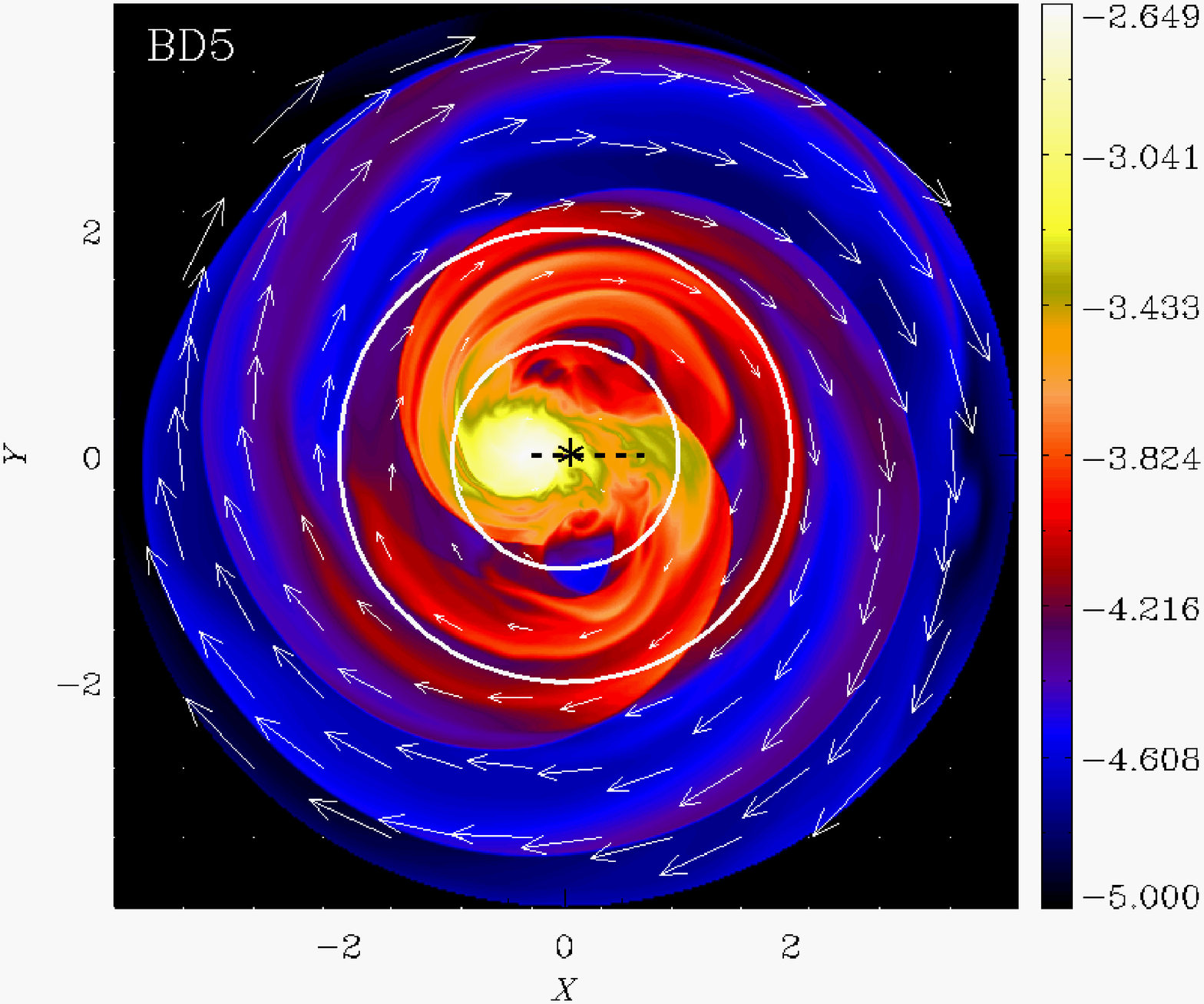}
\includegraphics[scale=0.3]{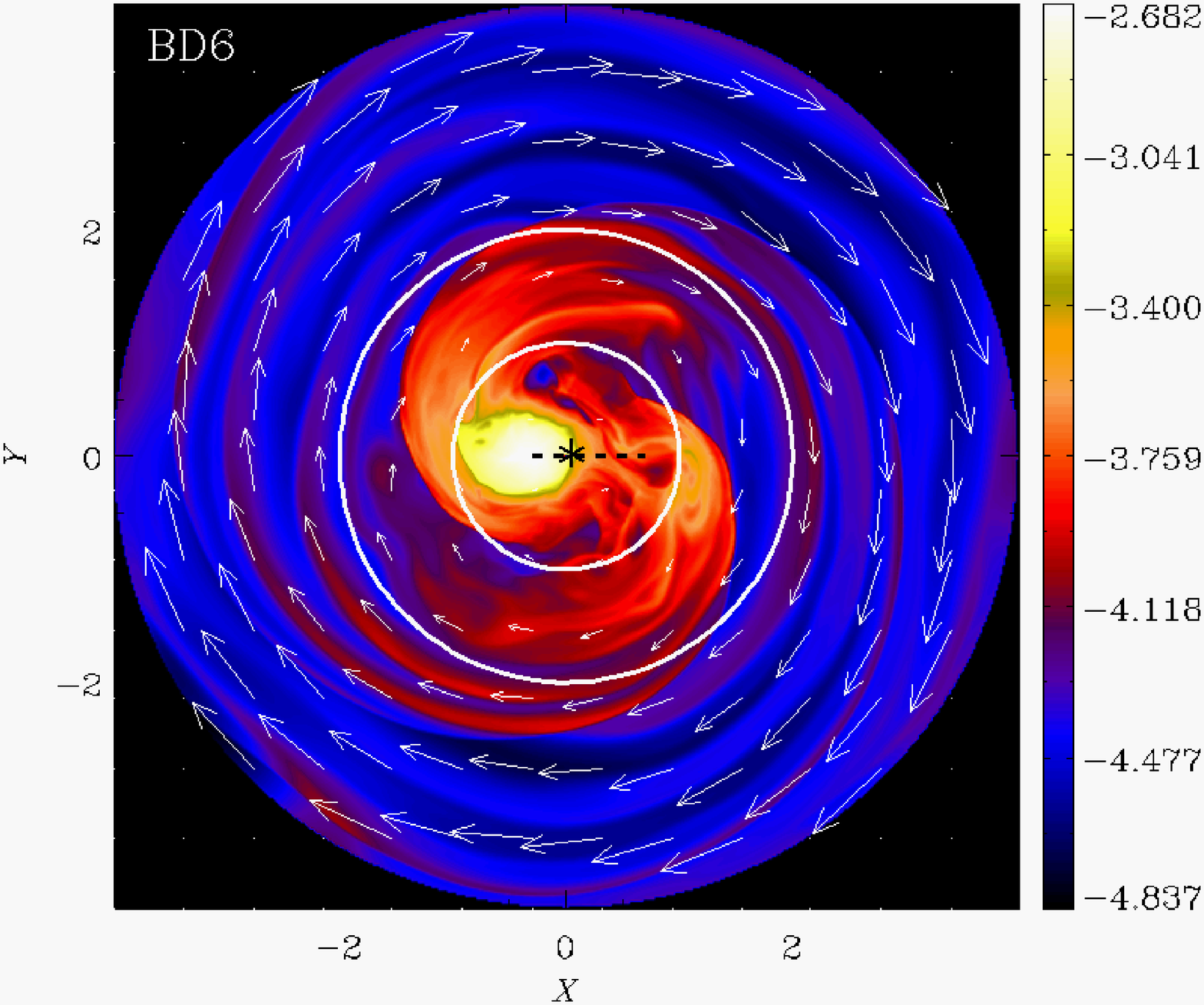}
\caption{
Same as Fig.~(\ref{rho_o1}) but for $\xi=0.3$.
}
\label{rho_o3} 
\end{figure*}

\begin{figure*}
\includegraphics[scale=0.3]{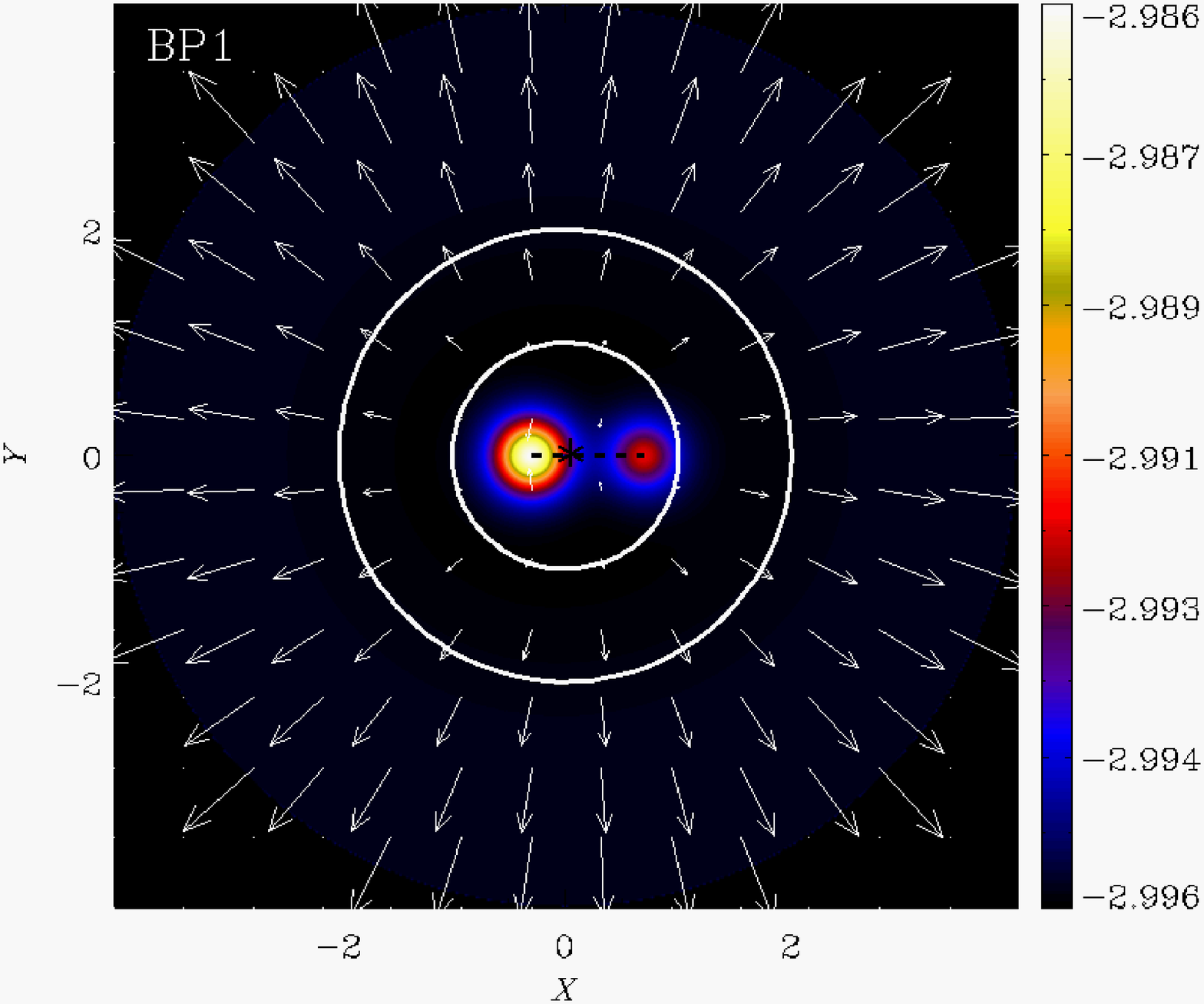}
\includegraphics[scale=0.3]{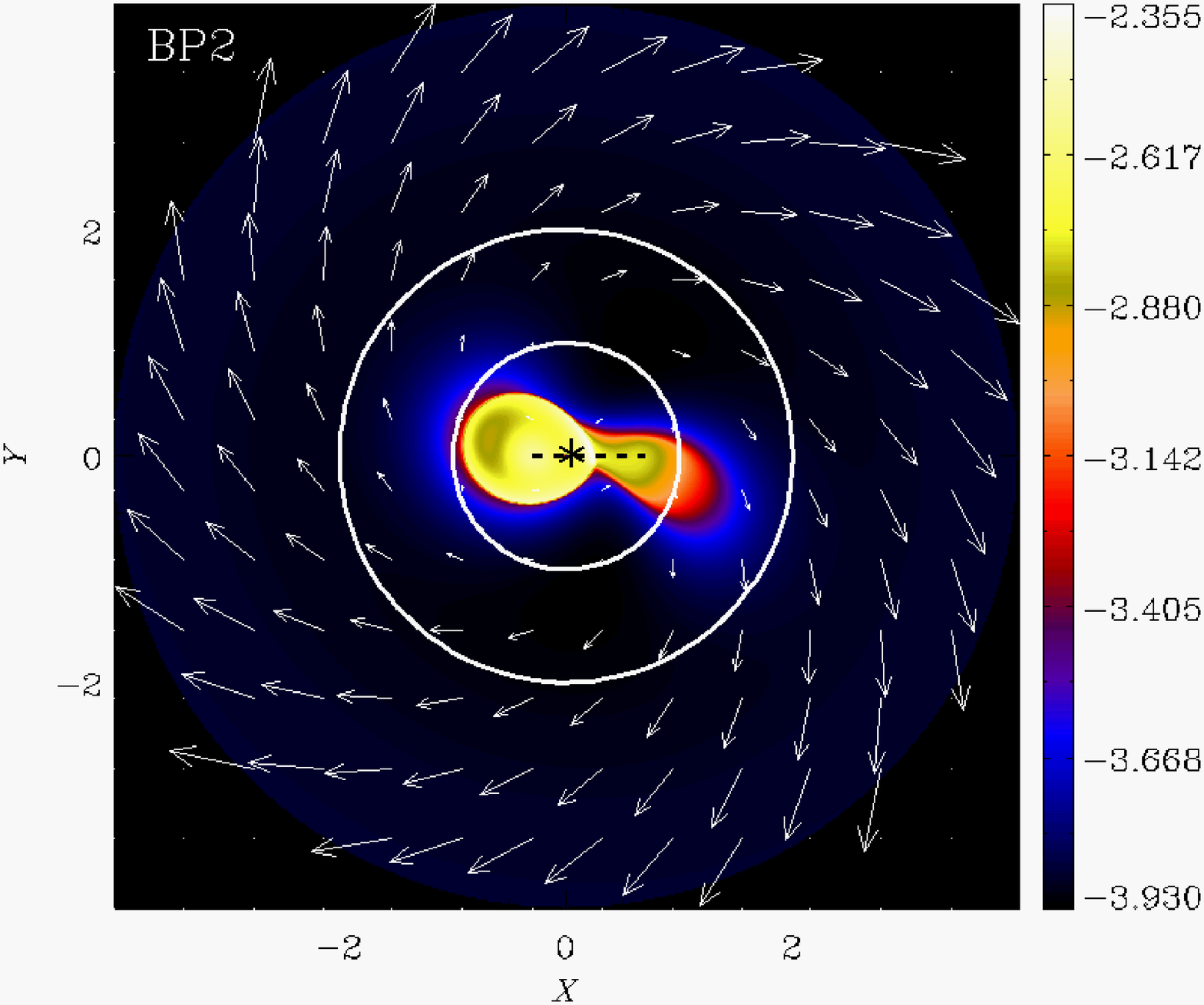}
\vskip0.01in
\includegraphics[scale=0.3]{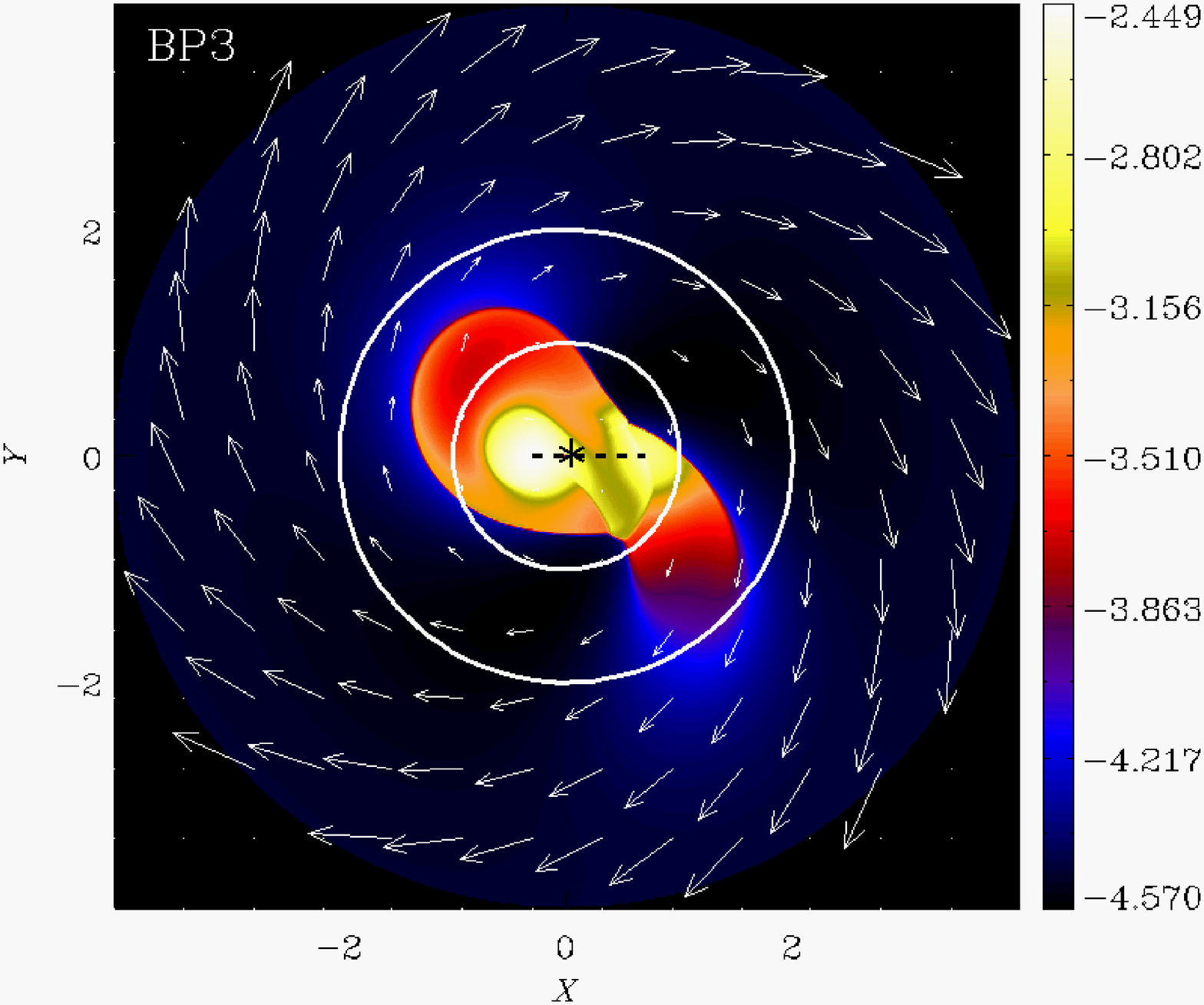}
\includegraphics[scale=0.3]{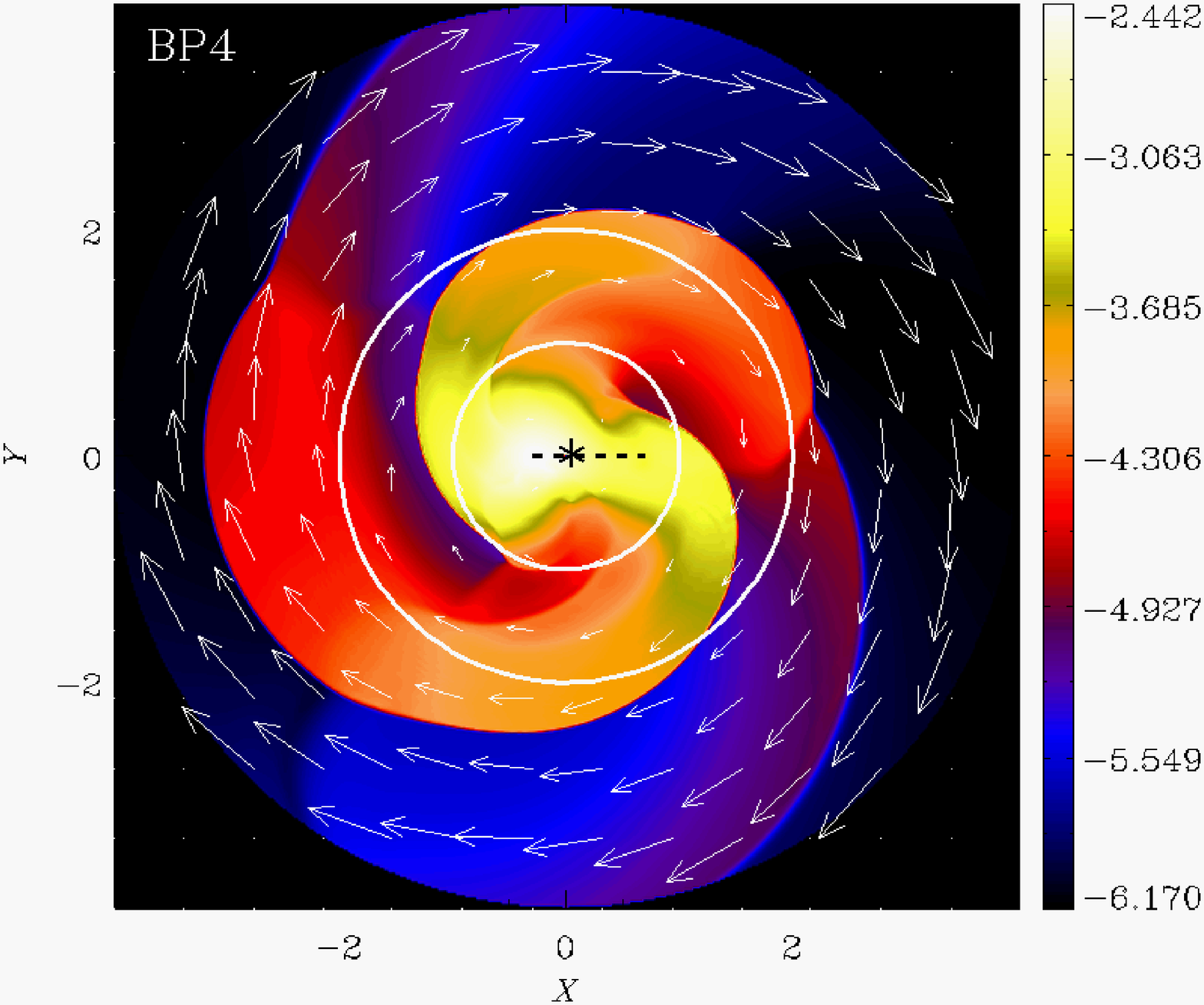}
\vskip0.01in
\includegraphics[scale=0.3]{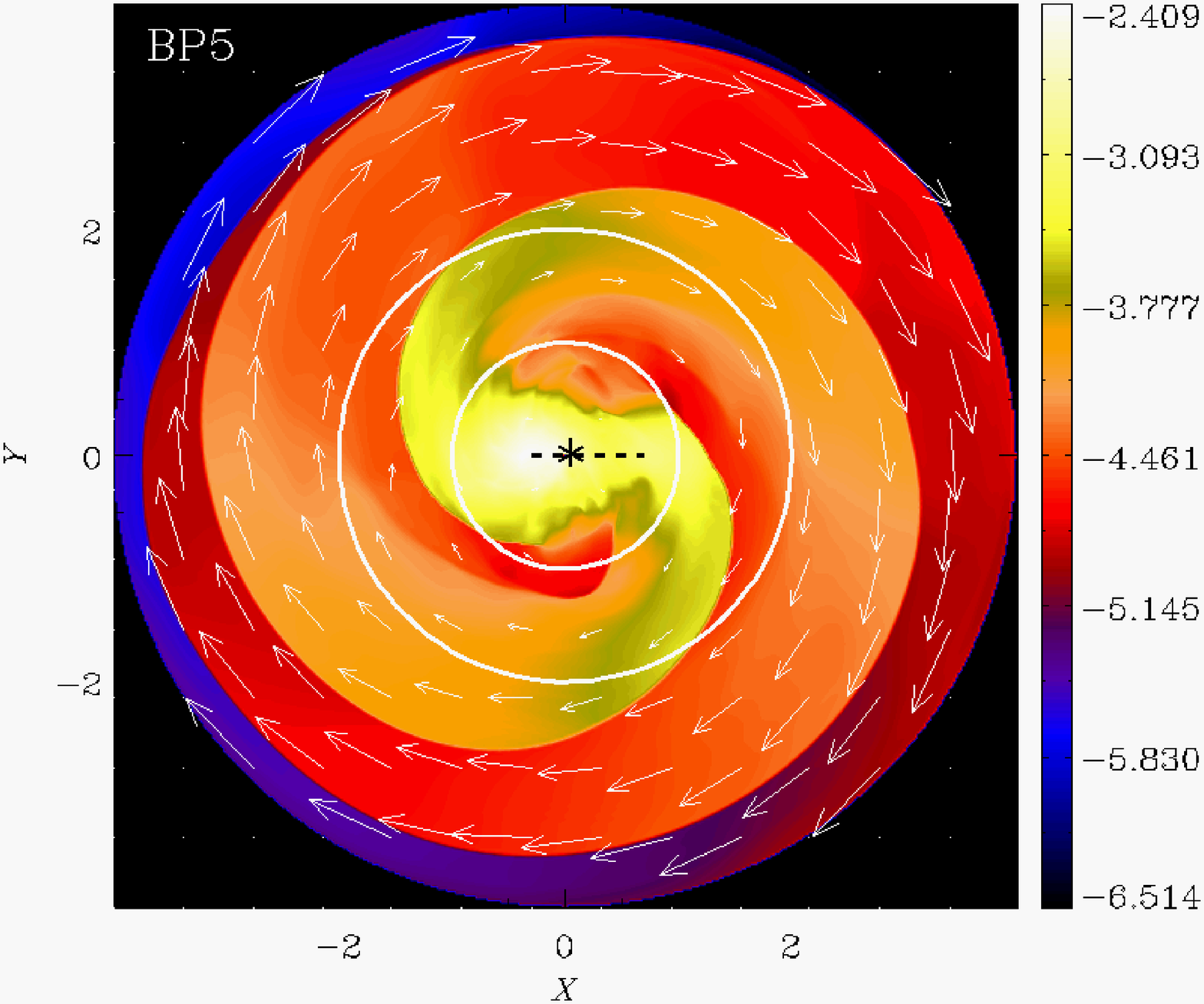}
\includegraphics[scale=0.3]{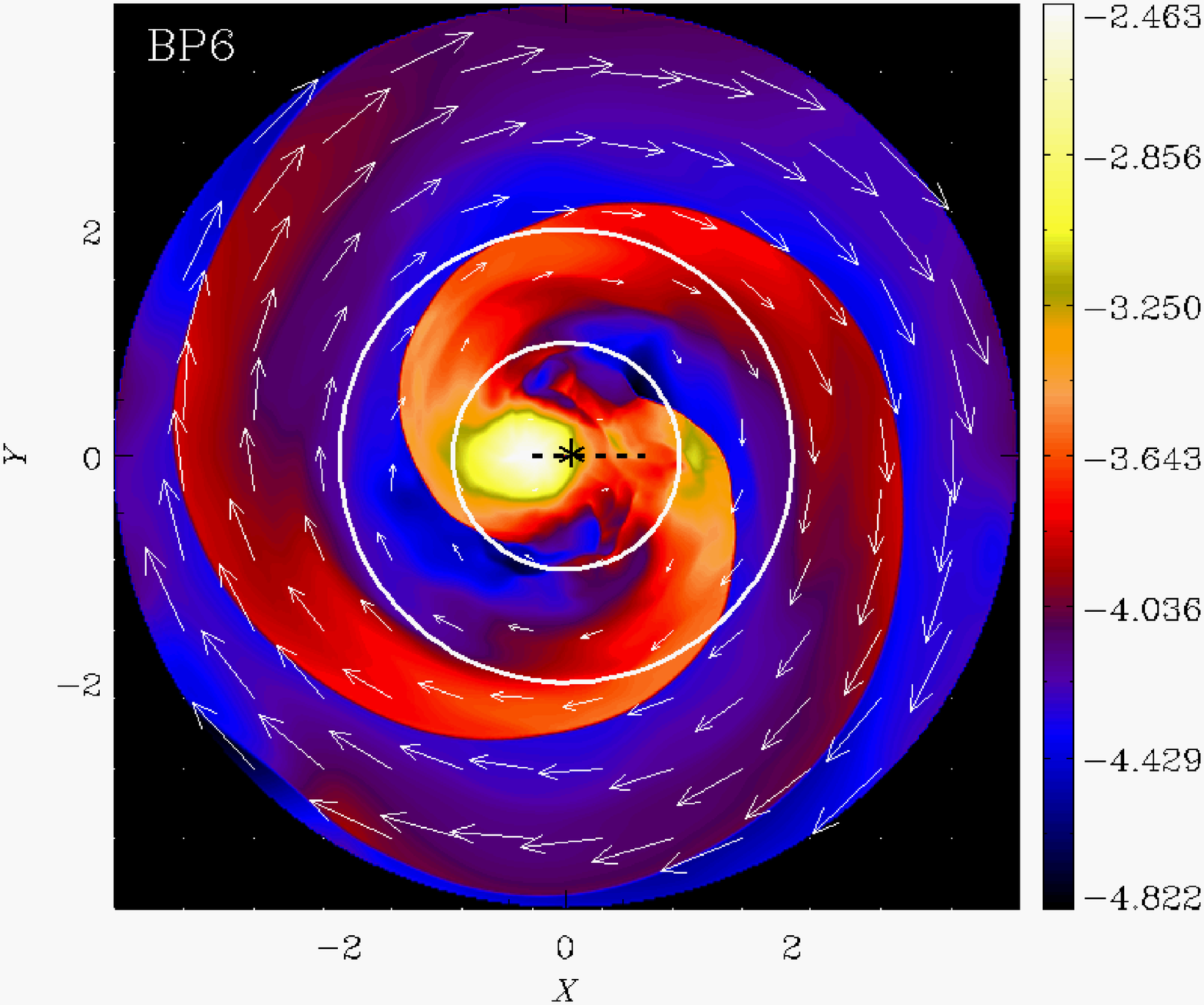}
\caption{
Same as Fig.~(\ref{prs_o1}) but for $\xi=0.3$.
}
\label{prs_o3} 
\end{figure*}

\subsection{Spiral nature of outflowing gas}

We present results from two different simulations with
$\xi=0.1$ (model~A) and $0.3$ (model~B),
and show the evolution of density/pressure maps with time as seen
in the corotating frame, which is the rest frame of the binary system.
The domain in $r$- and $\phi$- extends from $0$ to $4$ and
$0$ to $2\pi$, respectively, with ${\cal D}$ being the distance unit.
The grid resolutions were chosen to be $300\times 300$ for model~A
and $400\times 400$ for model~B, in $r\times \phi$ domain.
We reduced the total mass (${\mathscr M}$) of the binary by
factor $4$ in simulation with $\xi=0.3$ as compared to the one with
$\xi=0.1$. As noted before, ${\mathscr M}$ is the unit of mass in this
work.

In Figs~(\ref{rho_o1}) and (\ref{prs_o1}) we show, respectively,
the snapshots of the density and pressure, both in logarithmic
units ($\log_{10}\rho$ and $\log_{10}P$), for the model with $\xi=0.1$.
Time increases from panel AD1 (AP1) to AD6 (AP6) and we see the development
of the spiral pattern of the outflowing material. The evolution reaches a
steady state in later stages and thus we find that the structure of the
velocity field stops to evolve considerably. We note that the symmetry axis
of the spiral pattern is shifted from the rotation axis (which passes
through the center of mass of the binary) by distance $\xi$. The whole
pattern of density/pressure as shown in Fig~(\ref{rho_o1})/(\ref{prs_o1})
rotates with binary period for an inertial observer, and thus we
find periodic variation in the mass density along a given sightline.
To illustrate this, we have drawn two concentric circles, centered
around the common center of mass of the binary, showing the density/pressure
variations as one goes around the circles. The period of such
density modulations will be of the order of binary period.

Similarly we show snapshots of density and pressure for the model
with $\xi=0.3$ in Figs~(\ref{rho_o3}) and (\ref{prs_o3}), respectively.
As before we find that the stellar wind spirals outward and reaches
a steady state. The structure of the velocity field stays nearly the
same in late stages, which would be crucial to explain the observational
fact that the maser spot does not move in the sky.
Although our aim is to propose a physical model which can explain
some key qualitative features seen in the observations related to
maser intensity variations, we find that
a closer look of Figs~(\ref{rho_o1})--(\ref{prs_o3}) reveals
that the density at some spatial location along a particular
sightline might vary easily by a factor of about $10$ between its
maximum and minimum values. 

To make contact with physical units, we refer the reader to
Fig.~(\ref{mp}) where we show the relation between 
the total mass (${\mathscr M}$) and the time-period
(${\mathscr T}$) of the binary, corresponding to different values of
the binary separation (${\cal D}$); see also Eq.~(\ref{kep3-2}).
The mass, period and distance are expressed in terms of solar mass,
day and astronomical unit (AU), respectively.
If we let the distance unit in our simulations to be equal to
$1\, {\rm AU}$, the periods corresponding to total masses equal to
$1\, {\rm M}_{\odot}$ and $50\, {\rm M}_{\odot}$ will be about
$1\, {\rm year}$ and $50\, {\rm days}$, respectively.

\begin{figure*}
\includegraphics[scale=0.37]{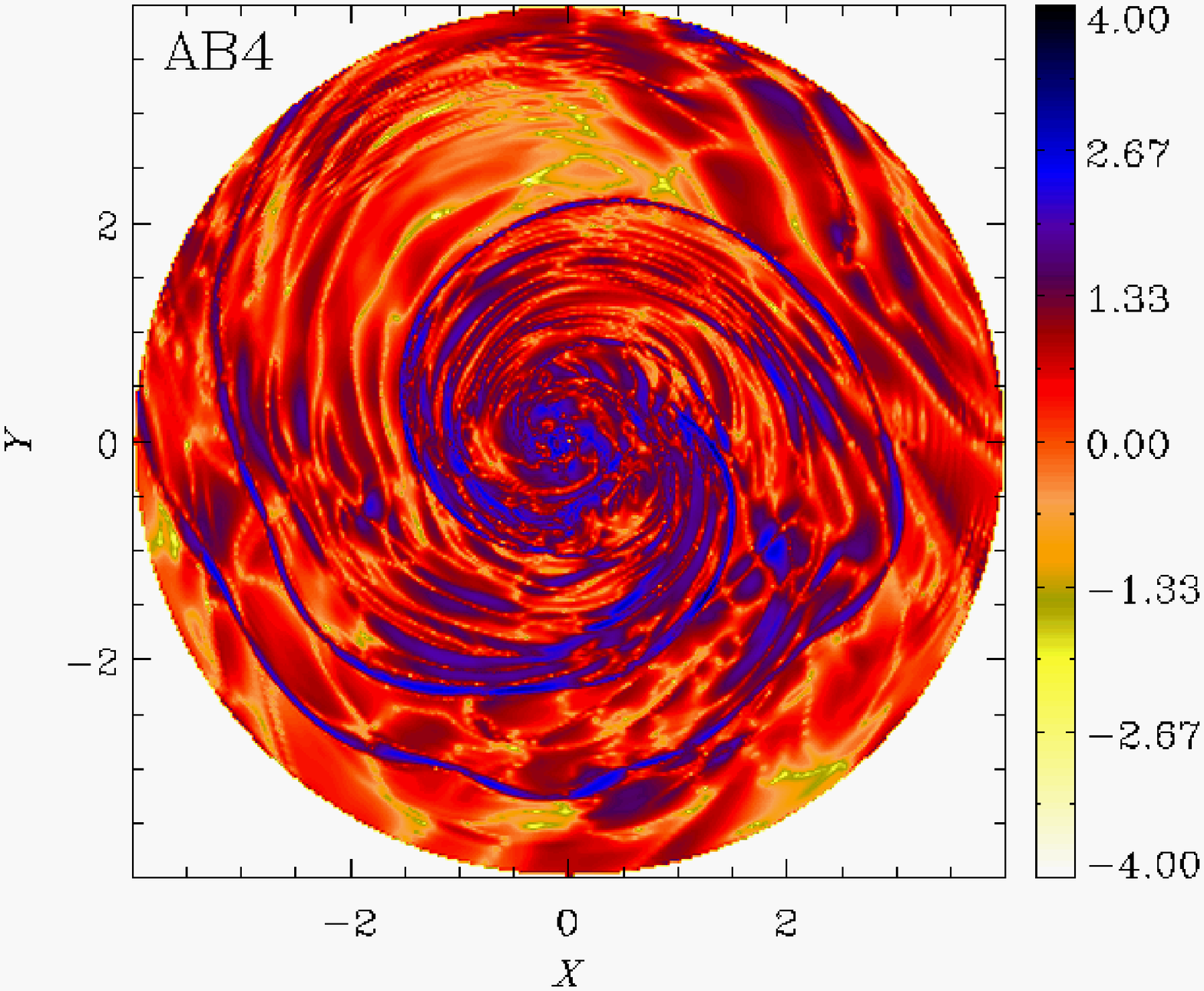}
\includegraphics[scale=0.37]{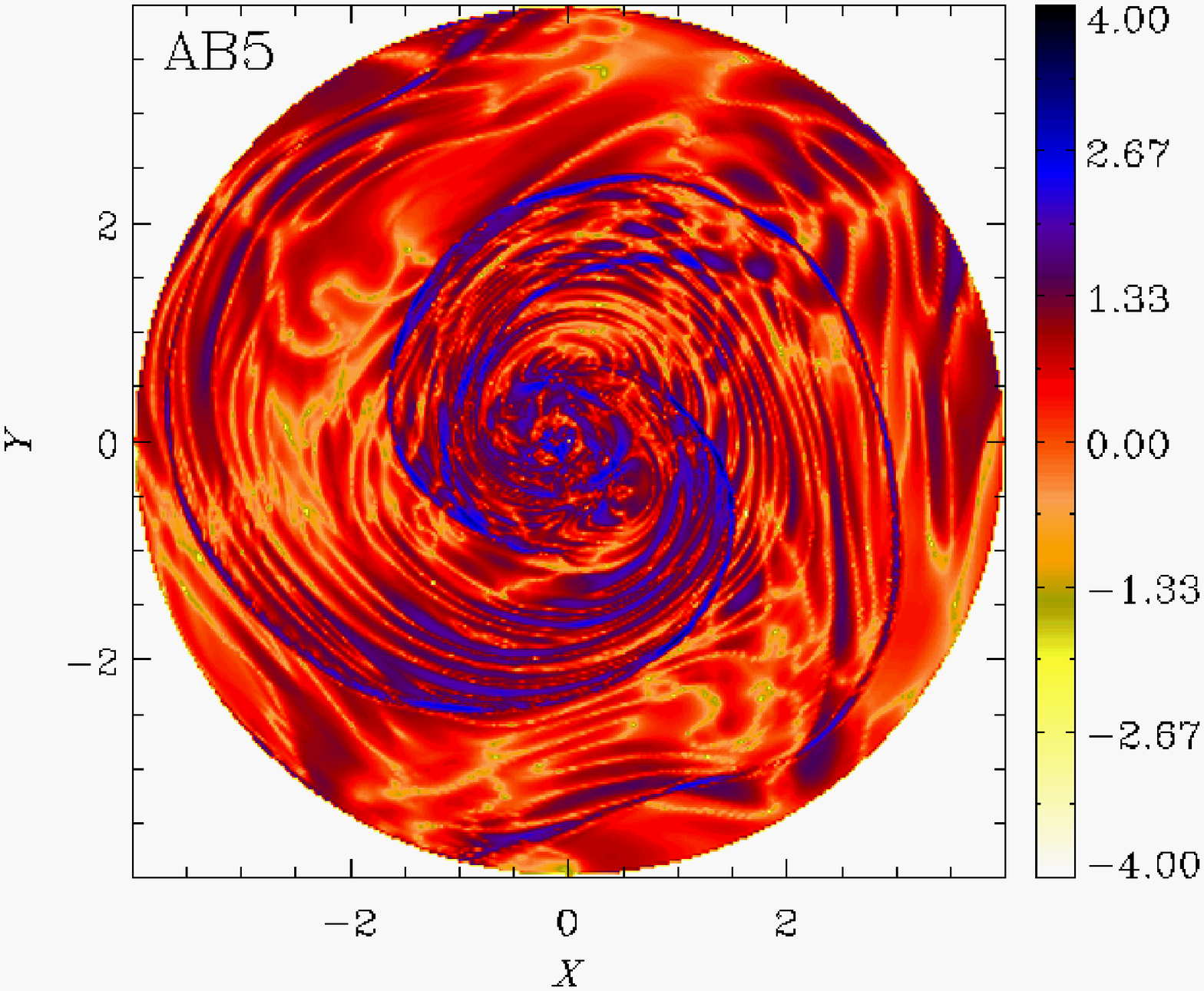}
\vskip0.01in
\includegraphics[scale=0.3]{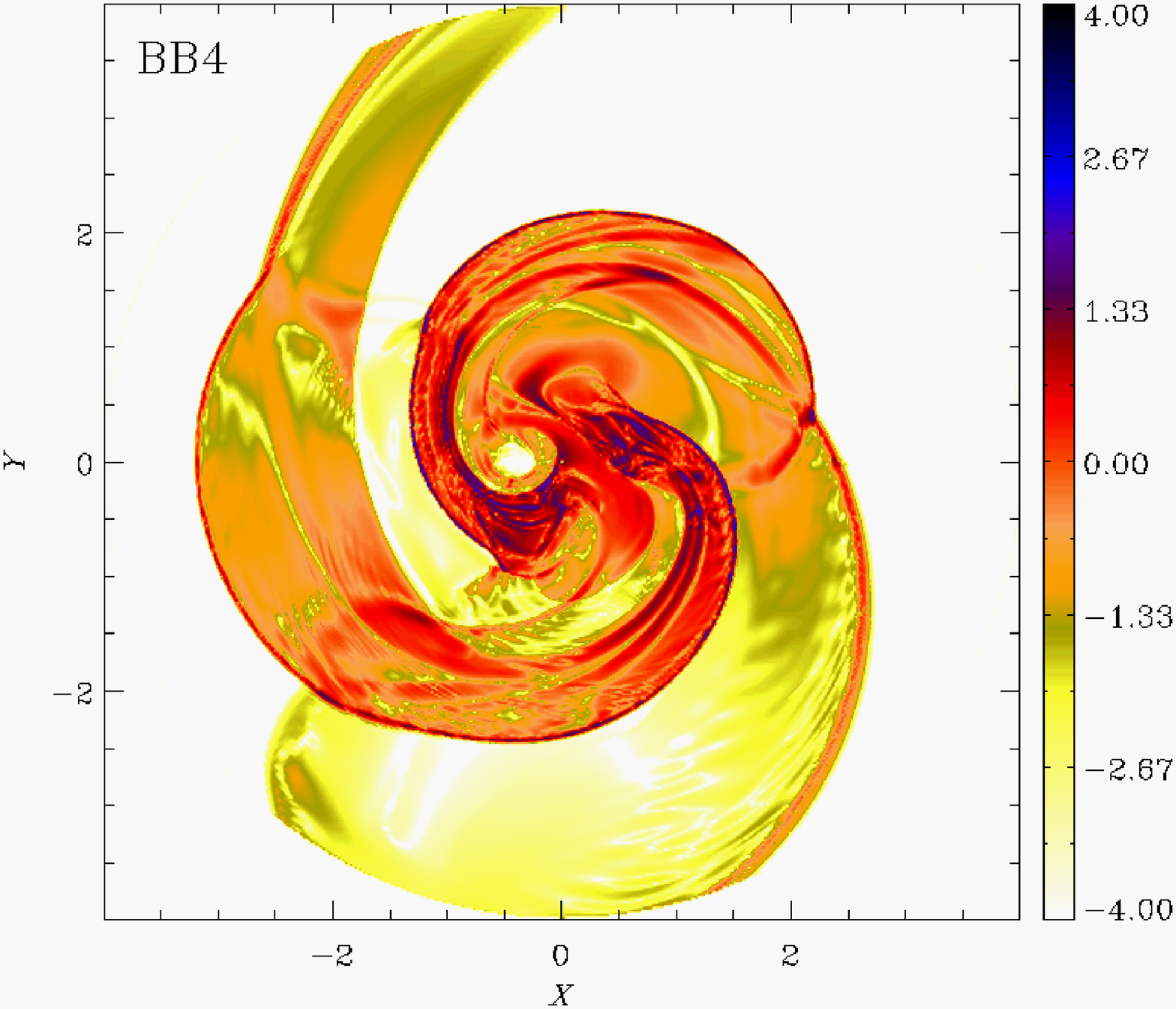}
\includegraphics[scale=0.3]{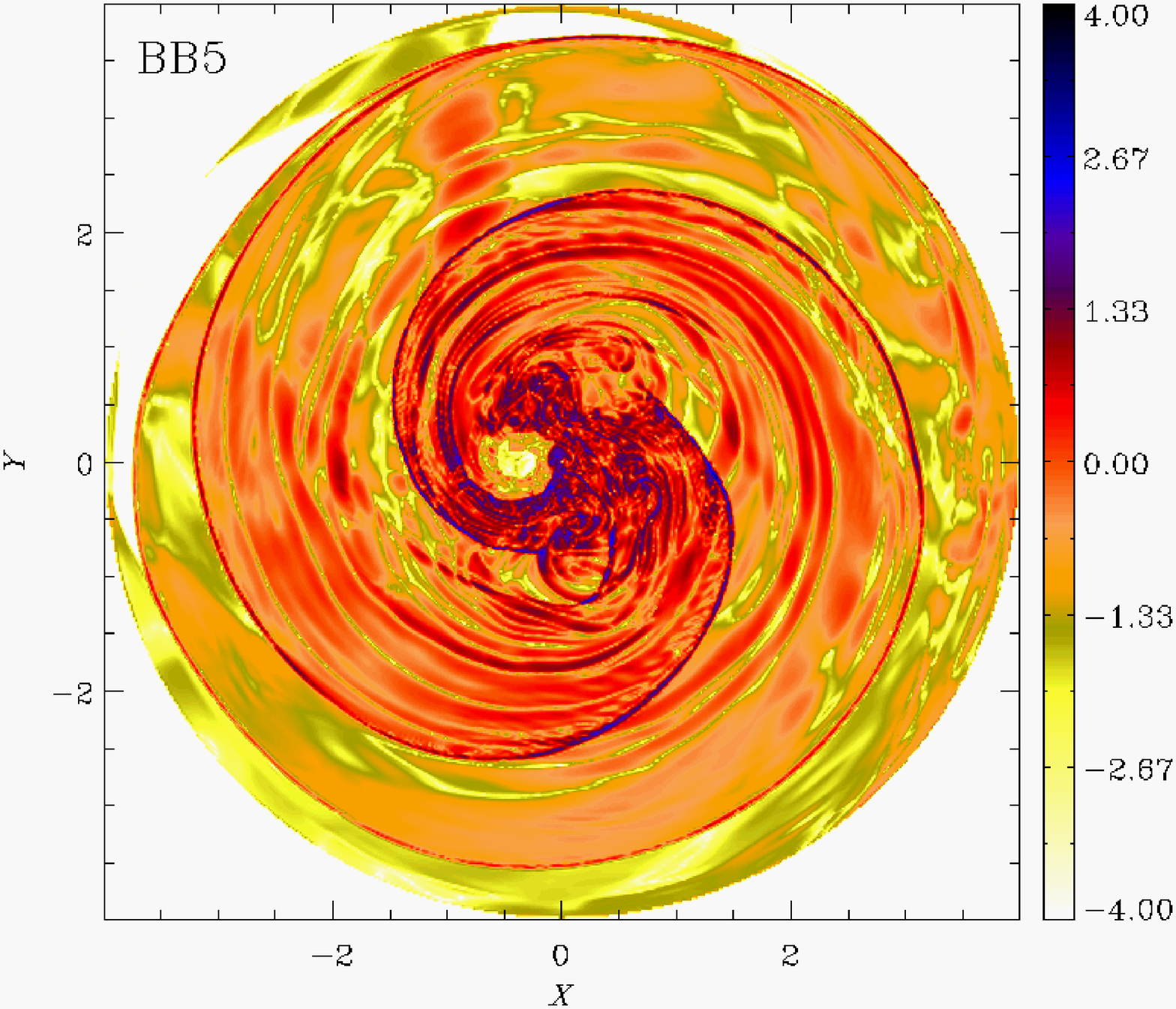}
\caption{
Snapshots of the quantity, $\log_{10}\vert \bfba \vert$ (shown by colours;
$\vert \bfba \vert \equiv \vert \beta_Z \vert$ in this setup),
for models~A (top panels; $\xi=0.1$) and B (bottom panels; $\xi=0.3$).
Left and right panels correspond to panels $4$ and $5$ of
Figs.~(\ref{rho_o1})-(\ref{prs_o3}).
}
\label{baro} 
\end{figure*}

\subsection{Baroclinicity}

We now turn to a completely different phenomena which will
have implications for the generation of the vorticity,
given by $\bfom=\bnabla\cross\bfv$. Given that the flow around
binaries in star-forming regions are expected to be highly
ionized, this is usually studied under standard magnetohydrodynamics
(MHD). Although the study of magnetic fields and the plasmas is beyond
the scope of the present paper, we, however, wish to remark that
the magnetic field, $\bfB$, could also evolve due to the presence of
baroclinicity, in the same model setup.
The mathematical form of the evolution
equations of the vorticity and the magnetic field are same;
these may be expressed as,
\begin{eqnarray}
\label{B}
\frac{\partial \bfB}{\partial t} &=&
\bnabla \cross (\bfv \cross \bfB) +
\frac{\bnabla P \cross \bnabla \rho_p}{\rho_p^2} +
\eta \nabla^2 \bfB \\[2ex]
\frac{\partial \bfom}{\partial t} &=&
\bnabla \cross (\bfv \cross \bfom) -
\frac{\bnabla P \cross \bnabla \rho}{\rho^2} +
\nu \nabla^2 \bfom
\label{om}
\end{eqnarray}
\noindent
where $\eta$, $\nu$ and $\rho_p$ are the magnetic diffusivity, the
kinematic viscosity and the plasma density, respectively.
The second term on the right hand side of Eq~(\ref{B})
(or \ref{om}) is known as the baroclinic term, which can lead
to the generation of $\bfB$ (or $\bfom$) even if the magnetic
field (or vorticity) was strictly zero to begin with.
The baroclinicity characterizing the misalignment
between the isocontours of the pressure and the density
may be expressed as,
\beq
\bfba \,=\, \frac{\bnabla P \cross \bnabla \rho}{\rho^2}
\label{beta}
\eeq

In the same model being discussed here, the baroclinic vector
field $\bfba$ is zero everywhere initially. It is instructive
to see if such a field could be generated at later stages of
evolving stellar wind in the Roche potential. To study this we
considered both the models~A and B, discussed
above, for which we have two dimensional data (in the plane of the binary)
of the density and the pressure fields at different instances
of times; few snapshots of density and pressure are shown
in Figs~(\ref{rho_o1})--(\ref{prs_o3}).
From this data, we compute $\bfba$ using Eq~(\ref{beta})
which points along $\pm \ez$ by definition, as both the vector fields,
$\bnabla P$ and $\bnabla \rho$, lie in the plane of the binary in our
2D setup. We find that $\beta$ changes sign in the binary plane, but its
magnitude is quite large, especially along the spirals.
In Fig.~(\ref{baro}) we show the snapshots of the quantity
$\log_{10}\vert \bfba \vert$ for
models~A (top panels) and B (bottom panels), in the late stages of the
simulations.

Thus in the same setup, we show that the baroclinicity develops
significantly, which can source the vorticity field
and also the magnetic field if the fluid is electrically conducting,
according to Eqs.~(\ref{B}) and (\ref{om}).
Similar conditions are expected to exist
in many parts of a galaxy, and therefore this provides yet another
possibility by which the seed magnetic fields could arise within the
galaxy. The spatial scales of variations of $\bfba$ are smaller
compared to the binary orbit and therefore this could source the seed
magnetic (or vorticity) fields at small scales.
Detailed investigations focussing on the magnetic fields, giving
quantitative estimates, are needed, and will be studied in a future
work.

\section{Maser variability (light-curve)}

Now we focus on the possible distribution of maser sources in the
medium around the binary system, as seen by the observer in the defined
geometry. The apparent (line-of-sight) velocity structure and the
column density of the molecular matter (for relevant species)
are amongst the key ingredients
governing the formation of astronomical masers.
Hydrodynamic simulations presented in \textsection~3 reveal
the flow structure and distribution of column density
in our model setup. We make use of these essential
informations and perform further numerical simulations to study
the masing action in such environment.
For simplicity, we have used a single-arm spiral pattern in these
simulations, instead of a two-arm spiral structure as suggested
from PLUTO simulations presented in the previous section.
We note that the patterns of field variables (mass density, pressure
and baroclinicity), shown in Figs~(\ref{rho_o1})--(\ref{baro}),
rotate with the angular speed ($\Omega$) of the binary, for
an observer in a fixed inertial frame. In this section we present
our results as would be seen by an inertial observer.

\begin{figure*}
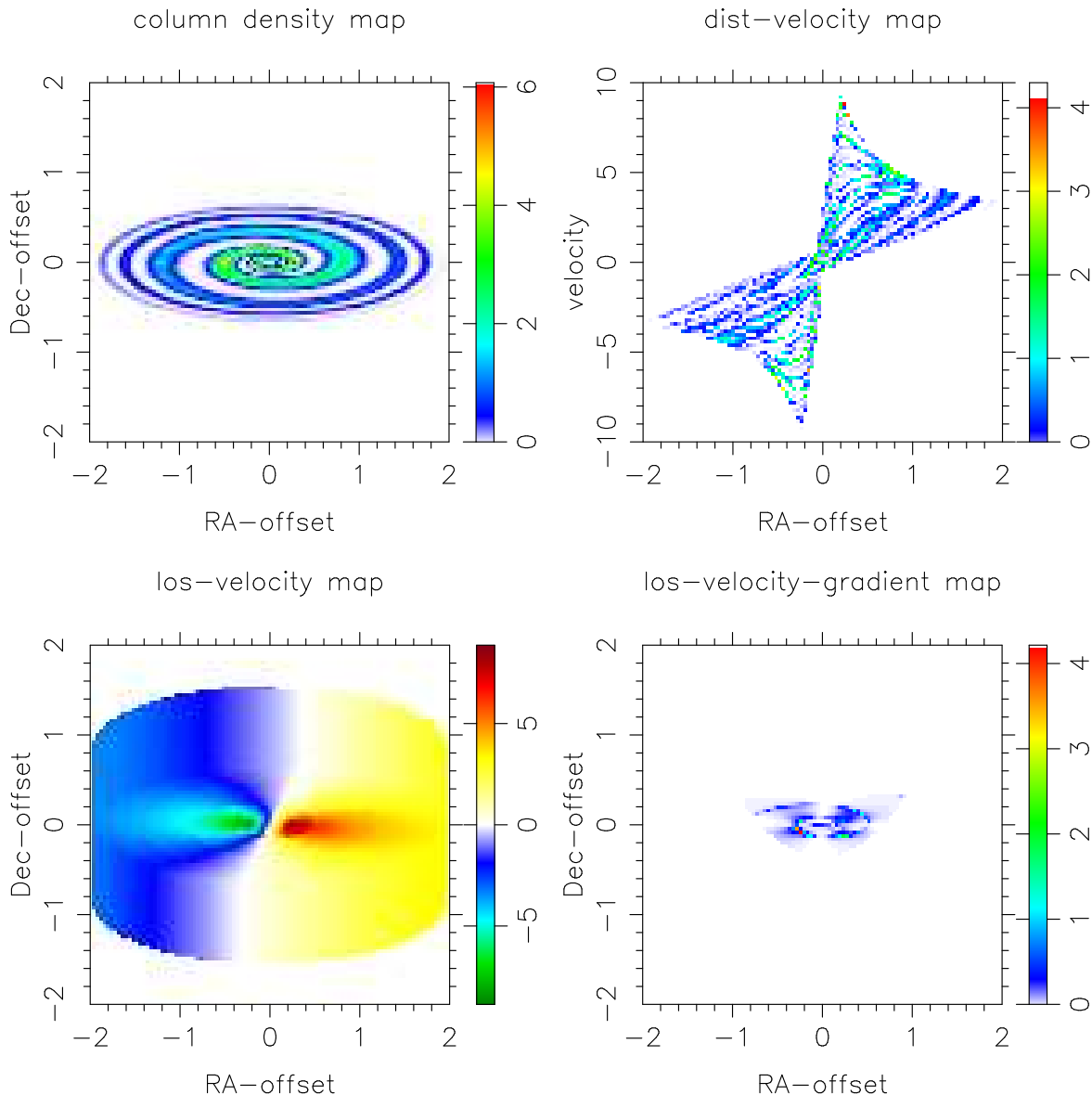

\includegraphics[scale=0.4,angle=-90]{spiral_model_plot_60_column_density}
\hskip0.1in
\includegraphics[scale=0.4,angle=-90]{spiral_model_plot_60_lv}
\vskip0.2in
\includegraphics[scale=0.4,angle=-90]{spiral_model_plot_60_los_velocity}
\hskip0.1in
\includegraphics[scale=0.4,angle=-90]{spiral_model_plot_60_velocity_gradient}
%\vskip0.05in
%\includegraphics[scale=0.4,angle=-90]{spiral_model_plot_60_intensity}
%\includegraphics[scale=0.4,angle=-90]{spiral_model_plot_60_transverse_x_velocity}
\caption{
Maps of (i) column density (top left); (ii) distance-velocity
(top right; colours: maser intensity); (iii) component of
velocity along the line of sight (bottom left);
and (iv) the gradient of velocity along sightline (bottom right),
in a fixed inertial frame,
shown for a binary system with angle of inclination $i_{\rm A}=60^{\circ}$.
Colors indicate the corresponding quantities on arbitrary scale.
}
\label{maps}
\end{figure*}

\begin{figure*}
\includegraphics[scale=0.68,angle=-90]{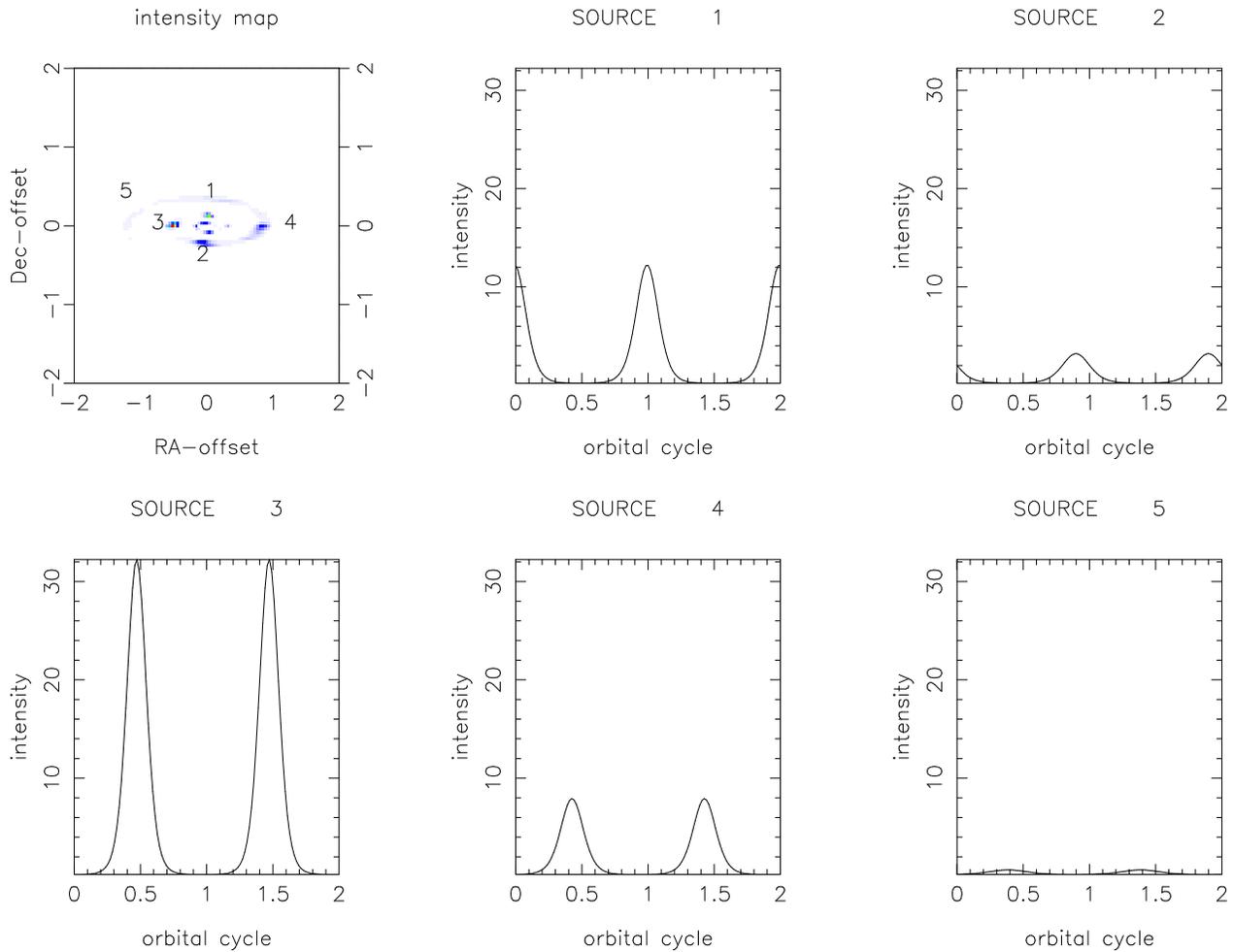}
\caption{
Maser intensity map around a binary system with an inclination angle
$i_{\rm A}=60^{\circ}$ shown in top left panel, where locations of
five masing spots are marked; seen from a fixed inertial frame.
Lightcurves of all these five sources are shown in the rest of the panels.
}
\label{light-curve}
\end{figure*}

Physical mechanisms governing the laws of maser formation in
cosmic settings are complex and have been studied extensively
(see e.g. \citet{GK72,Str74,Eli92}). A widely accepted model for
class II methanol masers is by \cite{SD94}.
Here our focus is on the conditions which can potentially cause
periodic variations in maser intensities, while also being able
to address the observational fact that the sky locations of the
maser spots stay fixed with time.
We do not intend to study the formation mechanisms of maser and
refer the reader to earlier works, some of them quoted above,
for further related studies.
For our purposes, it is sufficient to note that the intensity
of maser, ${\cal I} \propto \exp{(-\tau)}$, where $\tau$ is
optical depth (negative for amplification), which varies
along a particular sightline, and can be expressed in an average
sense as,
\beq
\tau \,\propto \, \frac{\Sigma}{\Delta v_{\parallel}}\,,
\label{I}
\eeq
\noindent
where $\Sigma$ is the mass column density of the medium and
$\Delta v_{\parallel}$ is the velocity spread along the chosen
line of sight.
Thus the sightlines with vanishing velocity gradients are ideal for masing
action. If the structure of the velocity field in the sky does not change,
then the masing spot is expected to stay at the same sky location.

Let $\bfR$ denote the location of the masing spot in the sky, seen from
the fixed coordinate frame $\overline{X}\,\overline{Y}\,\overline{Z}$.
If the density of the medium at $\bfR$ varies sinusoidally in time as,
$\rho(\bfR, t) \sim \rho_0 \sin{(\omega t + \varphi)} + \rho_{\rm f}$,
then, to lowest order in $\tau$, the maser intensity
will also show sinusoidal temporal modulation
with the same period ($=2\pi/\omega$), as may be seen from Eq~(\ref{I}).
Thus any temporal modulation in the column density
will result in maser intensity variation.
The quantities $\rho_0$, $\omega$, $\varphi$ and $\rho_{\rm f}$ denote,
respectively, the amplitude, angular frequency of modulation, the phase
and the floor density.
Given that the density maps shown in \textsection~3
rotate with angular speed ($\Omega$) of the binary, for an inertial observer
along a chosen line of sight, the observed density at $\bfR$ will
show modulations with the angular frequency ($\omega$) which is comparable
to $\Omega$, i.e., $\omega \sim {\cal O}(\Omega)$.
As we have modelled a single-arm spiral structure, for simplicity,
in this section (instead of two-arm spiral structure suggested from PLUTO
simulations), we note that the orbital periods will be overestimated by
factor two. We emphasize again that our aim here is to explore possible
physical conditions that can lead to observed periodic modulations in
maser intensities. One of the conclusions from our study may be stated as:
\emph{the binary period determines the period of maser intensity
variations.}

Binary orbits might be inclined at arbitrary angles with respect
to sightlines. The inclination angle ($i_{\rm A}$) may be defined as
the angle between normal to the binary plane and outwardly pointed
line of sight. We wish to illustrate the emission patterns when
the binary lies somewhere between the edge-on ($i_{\rm A}=90^{\circ}$)
and the face-on ($i_{\rm A}=0^{\circ}$) configurations. Here we consider
a binary with $i_{\rm A}=60^{\circ}$. We identify maser emitting regions
in our simulations and study the temporal evolutions of the maser
intensities from many different sources.

Figure~(\ref{maps}) shows maps of various physical quantities of interest
at some time and we briefly discuss these below:
\begin{enumerate}
\item [(i)] The column density shows single-arm spiral structure
where the density decreases as we move away from the central regions;
top left panel.
\item [(ii)] Remarkably, the distance-velocity diagram in top right
panel shows clearly the bipolar structure, as has been widely observed.
This is a natural outcome of our studies which suggests that
\emph{the observed bipolar nature of molecular outflows could also
arise due to presence of binary systems.}
\item [(iii)] Bottom left panel shows the map of the component of velocity
along the line of sight. It is useful to note that the structure of
the velocity field remains the same with time in this setup and therefore
the potential maser spots do not move in the sky, thus explaining this
observational fact.
\item [(iv)] The map of velocity gradient along the line of sight is shown in
bottom right panel. Low gradients in velocity determine the potential
sites where the masing action can take place, as may be seen from Eq~(\ref{I}).  
\end{enumerate}

This leads us to construct and monitor maser intensity maps as functions
of time. In Fig.~(\ref{light-curve}) we show a typical map of maser
intensity at some time (top left panel), where we mark the locations
of five different sources (masing spots). We intend to show the variations
of maser intensities as functions of time, or equivalently, as functions
of orbital cycle of the binary. Such plots of intensity variation,
commonly called light-curves, corresponding to all five sources are shown in other
panels of Fig.~(\ref{light-curve}). Some noteworthy properties are as
follows:
\begin{enumerate}
\item [(a)] We first notice that \emph{the maser intensities show periodic
temporal modulations.} These modulations are caused by the periodic
variation of column density, as discussed above, and the period is determined
by the binary period.
\item [(b)] Multiple masing spots could form around a single binary
system, and all these sources have same periods of modulation, as they
arise due to density variations, the period of which, in turn, is determined
by the binary system. However, the amplitudes of intensity variations
are different for different sources.
\item [(c)] Thus we argue that the observed periods of
such intensity variations trace the characteristics of the binary, i.e.,
the binary period.
This, together with some other independent estimates of either the mass
or separation of the binary, could be very useful in determining the
nature of binary systems, embedded in molecular clouds.
\item [(d)] Both, the arc-like and somewhat symmetric spots, are possible;
compare, e.g., spots `2' and `3' in Fig.~(\ref{light-curve}).
This might depend on the local physical conditions and the distances of the
spots from central regions.
Linear structures could indeed be reminiscent of Keplerian-like
velocity gradient, as has been conjectured by \cite{Nor98}.
\end{enumerate}

\section{Conclusions}

With an aim to provide physical mechanisms that could potentially
cause observed periodic variations in the maser intensities, we
studied the dynamics of stellar wind from one of the bodies in the
binary system. We find that the intensity variations are due to the
periodic variations in the local material density, where the period
is on the order of the binary period. The underlying velocity structure
is Keplerian-like and it remains frozen-in-time in the later stages.
We note that this non-evolving velocity structure is important for
the masing spots to stay at the same sky location, 
as has been revealed by the observations, 
and it is the maser intensity which varies periodically.
The wind appears to spiral outwards before settling into the faraway
Keplerian orbits, and naturally appears bipolar in the standard
position-velocity diagram.

The spherical symmetry of the stellar
wind is broken due to the influence of the other gravitating body.
This symmetry breaking very near to the star in the binary system
holds the key to establishing density inhomogenieties also far away
from the binary location where the conditions for masing action are
favourable. We first studied this probelm in the rotating frame in which
the binary components are at rest. To an inertial observer, such density
inhomogeneities in the rotating frame lead to a periodic density
variations along a chosen line of sight. Sightlines with nearly vanishing
velocity gradients along those directions being ideal for observing
masers show sinusoidal modulation in the material density, thus
causing the maser intensity also to show sinusoidal variation.
The structure of the velocity field being nearly frozen, which
is close to Keplerian, explains why these periodically varying maser
spots do not move. This mechanism also naturally gives a bipolar
appearance in the standard position-velocity diagram. We note that
the bipolar outflows are ubiquitously observed in nature and are
presumably thought to be associated with star forming regions in
molecular clouds. A binary origin may be responsible for some of
these bipolar outflows.

In the same setup, we also find that the baroclinicity develops and
it is mostly concentrated in the spiral form; see Fig.~(\ref{baro}).
This offers yet another scenario, which can lead to the
generation of seed vorticity and magnetic fields within the
galaxies, and might further affect the evolution of these
fields. More detailed investigations focussing primarily on
the magnetic fields will be presented in a future work.

\section*{Acknowledgments}
We thank Shuji Deguchi and Roy Booth for their interests and
encouragements during the IAU symposium 287 on Cosmic Masers,
held at Stellenbosch, South Africa. 
We are grateful to C. S. Shukre for discussions at an early stage
of this work. NKS thanks Mikhail Modestov for discussions on
baroclinicity, and Dipanjan Mukherjee for PLUTO-related issues.
We thankfully acknowledge the cluster facilities at RRI, IUCAA and
the Nordic High Performance Computing Center in Iceland, where the
computations were performed.

\label{lastpage}

\end{document}